\documentclass[aps,prb,twocolumn,showpacs,superscriptaddress]{revtex4-2}
\usepackage{amsfonts,mathrsfs,amsmath,amsthm,amssymb,bbold}
\usepackage[dvipdfmx]{graphicx}
\usepackage{bmpsize}
\usepackage{bm,mathrsfs}
\usepackage{lineno,hyperref}
\usepackage{amsmath}
\usepackage{color}
\usepackage{xcolor,colortbl} 
\usepackage{soul}
\usepackage{afterpage}
\usepackage{capt-of}
\usepackage{makecell}
\usepackage{ulem}
\usepackage{cancel}
\usepackage{siunitx}
\setcellgapes{4pt}
\newcolumntype{P}[1]{>{\centering\arraybackslash}p{#1}}

\begin{document}  
\title {\bf Spin and anomalous Hall effects emerging from topological degeneracy \\in Dirac fermion system \textbf{CuMnAs}} 

\author{Vu Thi Ngoc Huyen}
\affiliation{
Center for Computational Materials Science, Institute for Materials Research, Tohoku University, Sendai, Miyagi 980-8577, Japan}
\author{Yuki Yanagi}
\affiliation{
Center for Computational Materials Science, Institute for Materials Research, Tohoku University, Sendai, Miyagi 980-8577, Japan}
\author{Michi-To Suzuki}
\thanks{Electronic address: mts@tohoku.ac.jp}
\affiliation{
Center for Computational Materials Science, Institute for Materials Research, Tohoku University, Sendai, Miyagi 980-8577, Japan}
\affiliation{Center for Spintronics Research Network, Graduate School of Engineering Science,
Osaka University, Toyonaka, Osaka 560-8531, Japan}
\date{\today}

\begin{abstract}
Orthorhombic CuMnAs has been proposed as an antiferromagnetic semimetal hosting nodal line and Dirac points around the Fermi level. 
We investigate relations between the topological bands and transport phenomena, {\it i.e.} spin Hall effect and anomalous Hall effect, in orthorhombic CuMnAs with first-principles calculations combined with symmetry analysis of magnetic structures and of (spin) Berry curvature.
We show the nodal line gapped with spin-orbit coupling in CuMnAs dominantly generates large spin Hall conductivity in the ground state.
Although the magnetic symmetry in the ground state of CuMnAs forbids the finite anomalous Hall effect, applied magnetic fields produce a significant anomalous component of the Hall conductivity with the magnetic symmetry breaking. 
We identify that the dominant contribution to anomalous Hall components comes from further lifting of band degeneracy under external magnetic fields for the Bloch states generated with splitting of nodal lines by spin-orbit coupling near Fermi energy.
\end{abstract}
\maketitle
\section{Introduction}\label{secintro}
Transport phenomena such as anomalous Hall effect (AHE) and spin Hall effect (SHE) had been discovered to have a relation to the topological electronic band through Berry phase theory \cite{nagaosa,xiao}. These phenomena in antiferromagnetic (AFM) systems have essential advantages in comparison with ferromagnetic (FM) systems since there is no unexpected coupling at the interface and no perturbing stray field in the devices due to their magnetization \cite{Baltz_2018}.

Large AHE in antiferromagnets has been confirmed experimentally in Mn$_3$Sn \cite{2015mn3sn} and Mn$_3$Ge \cite{Kiyohara_2016} as predicted theoretically in the AFM states in Mn$_3$Ir \cite{2014Chen}, Mn$_3$Sn, and Mn$_3$Ge~\cite{2014mn3sn}.
Microscopic mechanism of large AHE has attracted for spintronics application in enhancing efficiency or replacing conventional materials. We investigated relations between the energy bands and the large AHE of metallic antiferromagnets Mn$_3A$N ($A$= Ni, Cu, Zn, Ga, Ge, Pd, In, Sn, Ir, Pt) in earlier work~\cite{mn3an} and identified that the Berry curvature whose size is small but spread widely around the Fermi surfaces in Brillouin zone (BZ) dominantly contribute to the anomalous Hall conductivity (AHC) rather than the Berry curvature with extremely large intensity within the local regions around the Weyl points.
The presence of large Fermi surfaces thus makes it difficult to distinguish the specific local contribution of topological bands, such as nodal lines and Weyl/Dirac points, 
to the transport quantities in these metallic magnetic compounds.
Therefore, further investigation for semi-metallic magnetic compounds that have only small Fermi surfaces is useful to understand relation between the topological bands and transport properties in complex magnetic systems as was done for large anomalous Nernst effect of magnetic Weyl semimetal in recent studies of Co$_2$MnGa and Co$_3$Sn$_2$S$_2$ \cite{Sakai_co2mnga_2018,Guin_co3sn2s2_2019,Minami_2020, Yanagi_2020}.

The SHE was theoretically investigated in non-magnetic compounds as well as magnetic transition metals~\cite{Tanaka_2008,Freimuth} and observed experimentally in metallic AFM compounds PtMn, IrMn, PdMn, and FeMn with highlighting the role of spin-orbit coupling (SOC) to the SHE \cite{Zhang}.
The anisotropic property of SHE, {\it i.e.} the difference in tensor components of the spin Hall conductivity, in PtMn, IrMn, PdMn, FeMn \cite{Zhang}, and Zr$XY$ ($X$=Si, Ge, $Y$= S, Se, Te) \cite{turnable} leads to a possibility to tune large SHE by exploiting the change in the direction of electric fields. 
Besides the electric fields, magnetic fields can control both the SHE and AHE by symmetry breaking. Therefore, investigation of the origin of SHE and AHE with and without an external magnetic field can pave the way to manipulate the transport phenomena in AFM systems.

The AFM orthorhombic CuMnAs was recently found to be a good candidate for exploring spintronic applications with specific topological features such as topological metal-insulator transition and topological anisotropic magnetoresistance with the presence of Dirac fermions \cite{cumnas3,cumnas4,cumnas5,cumnas6}. 
We further expect that the semimetallic magnetic ground state of the CuMnAs provides a deeper understanding of the possible contribution from the Bloch states related to topologically protected degeneracy to transport phenomena with its intensive investigation.
In this paper, we adopt the semimetallic AFM states of CuMnAs as a platform to investigate relations between topological bands, such as Dirac/Weyl points and nodal lines, and transport quantities of SHE and AHE. 

The high-symmetric collinear AFM order of CuMnAs forbids AHC from being finite due to the preserved symmetry of the combined operation of spacial inversion $P$ and time-reversal $T$. However, as we discuss later, applied magnetic fields can induce anomalous components of the Hall conductivity by breaking the $TP$ symmetry for the local structure around the nodal line of the electronic bands in BZ.

The organization of the paper is as follows.
Section~\ref{method} presents the method to perform the first-principles calculations and to calculate the spin Hall conductivity (SHC) and anomalous Hall conductivity (AHC). 
Section~\ref{syms} discusses the symmetry aspects of AHE and SHE in CuMnAs with general relation of spin Berry curvature.
Then, section~\ref{results} gives computed results about
stability of AFM structures and their electronic property, along with discussions of topological bands such as Dirac nodal lines in relation to the AHE and SHE, including effects of SOC and of external magnetic fields.
Finally, Sec.~\ref{conclusion} contains a summary of this work.
\section{Method}\label{method}
To investigate relations between specific band structures and transport properties, we study AHE and SHE with momentum-space Berry phase theory \cite{berry1984,velocity3}. The AHC $\sigma^{\mathrm{A}}_{\alpha\beta}$ and SHC $\sigma^{\mathrm{S},\gamma}_{\alpha\beta}$ are evaluated following the Kubo formula \cite{ahc, ahc4}:
\begin{equation}
\label{equ:ahc}
\begin{split}
\sigma^{\mathrm{A}}_{\alpha\beta}=-\frac{e^2}{\hbar}\int\frac{d\textbf{\textit{k}}}{(2\pi)^3}\sum_nf_n(\textbf{\textit{k}})\Omega^{\mathrm{A}}_{n,\alpha\beta}(\textbf{\textit{k}}),
\end{split}
\end{equation}
\begin{equation}
\label{equ:shc}
\begin{split}
\sigma^{\mathrm{S},\gamma}_{\alpha\beta}=\frac{\hbar}{2e}\cdot \frac{e^2}{\hbar}\int\frac{d\textbf{\textit{k}}}{(2\pi)^3}\sum_nf_n(\textbf{\textit{k}})\Omega^{\mathrm{S},\gamma}_{n,\alpha\beta}(\textbf{\textit{k}}),
\end{split}
\end{equation}
where $n$ is band index, $\alpha,\beta, \gamma= x,y,z$ ($\alpha\neq\beta$ for the AHC components), $f_{n\textbf{\textit{k}}}=\theta(\mu-\epsilon_{n\textbf{\textit{k}}})$ is the occupation factor determined from the eigenvalue of the Bloch states $\epsilon_{n\textbf{\textit{k}}}$ and the Fermi energy $\mu$. 
$\Omega^{\mathrm{A}}_{n,\alpha\beta}(\textbf{\textit{k}})$ and $\Omega^{\mathrm{S},\gamma}_{n,\alpha\beta}(\textbf{\textit{k}})$ are Berry curvature and spin Berry curvature, respectively, defined as:
\begin{equation}
\label{equ:berry}
\begin{split}
\Omega^{\mathrm{A}}_{n,\alpha\beta}(\textbf{\textit{k}})= -2 {\hbar}^2 \mathrm{Im} \sum_{m\neq n}\frac{v_{nm,\alpha}(\textbf{\textit{k}})v_{mn,\beta}(\textbf{\textit{k}})}{[\epsilon_m(\textbf{\textit{k}})-\epsilon_n(\textbf{\textit{k}})]^2},
\end{split}
\end{equation}
\begin{equation}
\label{equ:spinberry}
\begin{split}
\Omega^{\mathrm{S},\gamma}_{n,\alpha\beta}(\textbf{\textit{k}})= -\frac{4 \hbar}{e} \mathrm{Im} \sum_{m\neq n}\frac{j^{\gamma}_{nm,\alpha}(\textbf{\textit{k}})j_{mn,\beta}(\textbf{\textit{k}})}{[\epsilon_m(\textbf{\textit{k}})-\epsilon_n(\textbf{\textit{k}})]^2},
\end{split}
\end{equation}
where velocity operator is calculated with the periodic part of the Bloch states, $u_{n\textbf{\textit{k}}}$, as:
\begin{equation}
\label{equ:velocity}
\begin{split}
v_{nm,\alpha}(\textbf{\textit{k}})=\frac{1}{\hbar}\left \langle u_{n}(\textbf{\textit{k}})\left|\frac{\partial \hat{H}(\textbf{\textit{k}})}{\partial k_{\alpha}}\right | u_{m}(\textbf{\textit{k}}) \right \rangle,
\end{split}
\end{equation}
with $\hat{H}(\bm{k})=e^{-i\bm{k}\cdot\bm{r}}\hat{H}e^{i\bm{k}\cdot\bm{r}}$, 
$j_{\beta} = e v_{\beta}$ is charge current operator, and $\hat{j}^{\gamma}_{\alpha}=\frac{1}{2}\{\hat{s}_{\gamma},\hat{v}_{\alpha}\}$ is spin-current operator with spin operator $s_\gamma$.
The (spin) Berry curvature is expected to increase divergently around band crossing point due to the denominator of Eq.~(\ref{equ:berry}) [Eq.~(\ref{equ:spinberry})]. As a result, the accurate calculation of AHE (SHE) usually require dense $k$-mesh for the BZ integration of the (spin) Berry curvature in Eq.~(\ref{equ:ahc}) [Eq.~(\ref{equ:shc})].

The first-principles calculations for magnetic states without an external magnetic field are performed by QUANTUM ESPRESSO package \cite{QE}. 
The generalized gradient approximation (GGA) in the parameterization of Perdew, Burke, and Ernzerhof~\cite{GGA-PBE} is used for the exchange-correlation functional and the pseudopotentials in the projector augmented-wave method~\cite{paw_original,paw} are generated by PSLIBRARY~\cite{pslibrary}. We choose kinetic cut-off energies 50 Ry and 400 Ry for the plane-wave basis set and charge density, respectively. Lattice constants were taken from experimental values~\cite{cumnas6}, which are $a=6.572 {\rm \AA}, b=3.861 {\rm \AA}$, and $c=7.305 {\rm \AA}$; atomic potions are fully relaxed with keeping the lattice constants. 
A $k$-mesh 9$\times$15$\times$9 is utilized to sample the first BZ with Methfessel-Paxton smearing width of 0.005 Ry to get the Fermi level. The SHC in the absence of an external magnetic field is evaluated with PAOFLOW package~\cite{paoflow}.

The ELK code \cite{elk} and Wannier90 \cite{w90} are used to investigate the magnetic field dependence of magnetization, AHE, and SHE under the external magnetic fields.
The $3p, 3d,$ and $4s$ orbitals of Cu, the $3s, 3p, 3d,$ and $4s$ orbitals of Mn, the $3d, 4s,$ and $4p$ orbitals of As are treated as band states.
The $4s,3d$ orbitals for Cu and Mn atoms and $4s, 4p$ orbitals for As atoms are included for the Wannier interpolation scheme using Wannier90 to construct the realistic tight-binding models obtained from the first-principles band structures \cite{ahc}. 
Most of the AHC and SHC values reach the convergence within a few percent under the evaluation with the uniform $k$-point mesh of 180$\times$280$\times$180 and the adaptive $k$-mesh refinement \cite{adap1,adap2} of 5$\times$5$\times$5 for the absolute values of the (spin) Berry curvature larger than 100$\mathrm{\AA^2}$. Some calculations, which need a larger $k$-mesh, will be noticed in Sec.~\ref{results}. 
\section{Symmetry aspects}\label{syms}
\begin{figure} 
\centering 
\includegraphics[width=7.5cm]{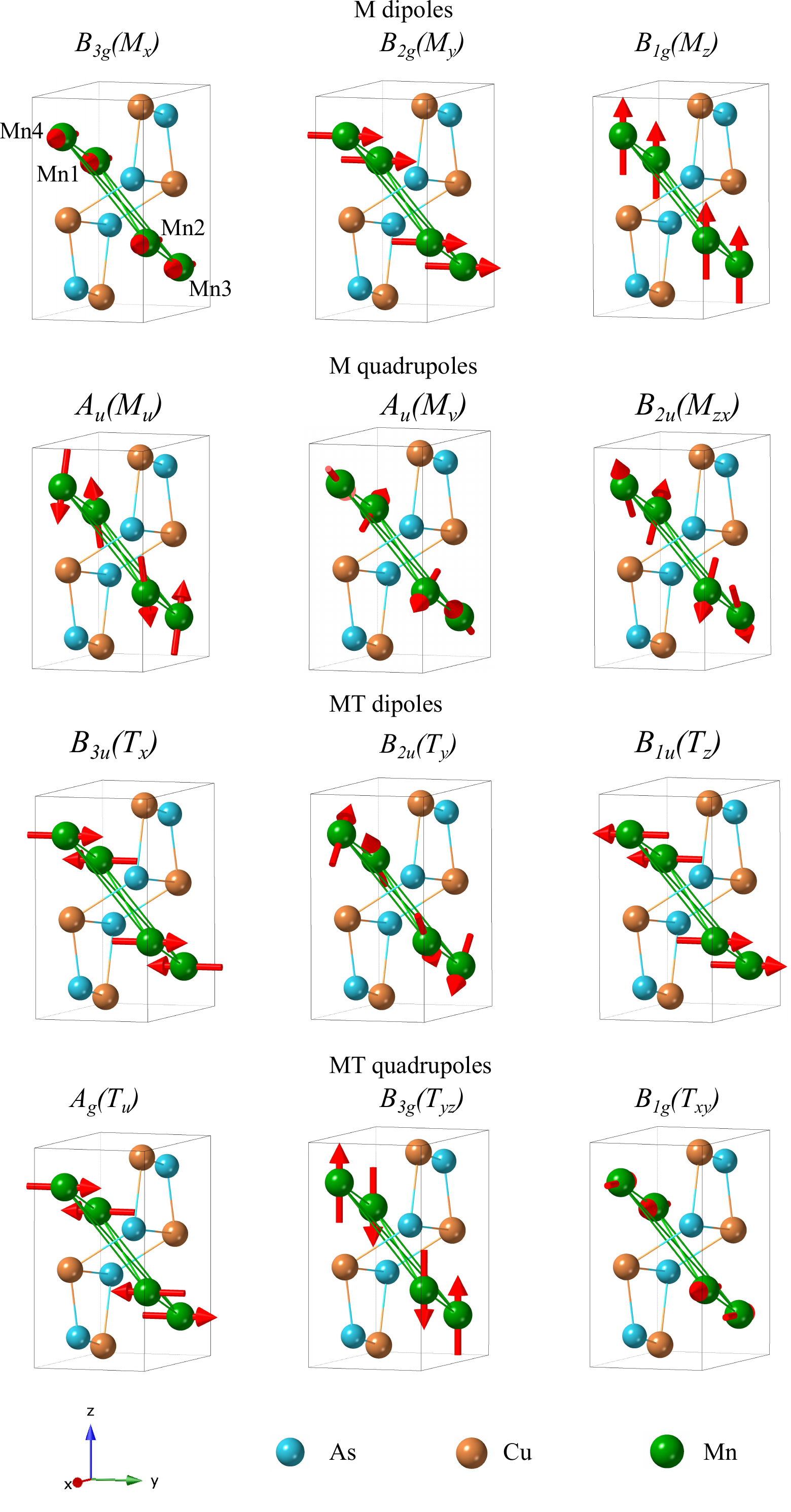}\\
\captionof{figure}{Energetically inequivalent magnetic structures of CuMnAs classified according to the multipole moments as magnetic (M) dipoles, magnetic (M) quadrupoles, magnetic toroidal (MT) dipoles, and magnetic toroidal (MT) quadrupoles following the Ref.~\onlinecite{cmpgeneration}. The arrows on Mn atoms indicate the magnetic moments. Four manganese atoms are marked as Mn1, Mn2, Mn3, and Mn4 for $B_{1g}(M_z)$, on which the magnetic alignments are listed in Table~\ref{tab:orthonormal}.}
\label{fig:orth}
\end{figure}
\begin{table*}
\captionof{table}{Classification of the magnetic structures with the ordering vector ${\bm q} = \bm{0}$ in CuMnAs according to the symmetry-adapted multipole~\cite{cmpgeneration} as well as the irreducible representations (IREPs). Magnetic moments alignments for four manganese atoms are listed along with magnetic space group (MSG), magnetic point group (MPG), and the AHC tensor components (AHC) that can be finite under the magnetic point groups which are also shown in \cite{Grimmer_1993,seemann}.}
\label{tab:orthonormal}
\begin{tabular}{P{1.6cm}P{1.8cm}P{7.8cm}P{1.8cm}P{1.8cm}P{1.8cm}} 
\hline
\hline
 IREP & Multipole & Magnetic moments alignment & MSG & MPG & AHC\\
 \hline
 $B_{3g}$& $M_x$ & $\left(\begin{array}{rrrr} \mathrm{Mn1:} & 1.0 & 0.0 & 0.0 \\ \mathrm{Mn2:} & 1.0 & 0.0 & 0.0 \\ \mathrm{Mn3:} & 1.0 & 0.0 & 0.0 \\ \mathrm{Mn4:} & 1.0 & 0.0 & 0.0  \end{array}\right)$ &$Pnm'a'$& $m'm'm$&$\sigma_{yz}$\\
$B_{2g}$& $M_y$ &$\left(\begin{array}{rrrr} \mathrm{Mn1:} & 0.0 & 1.0 &0.0 \\ \mathrm{Mn2:}& 0.0 & 1.0 & 0.0 \\ \mathrm{Mn3:}& 0.0 & 1.0 & 0.0 \\ \mathrm{Mn4:} &0.0 & 1.0 & 0.0 \end{array}\right)$& $Pn'ma'$&$m'm'm$&$\sigma_{zx}$\\
$B_{1g}$& $M_z$ &$\left(\begin{array}{rrrr} \mathrm{Mn1:}& 0.0 & 0.0 & 1.0 \\ \mathrm{Mn2:}& 0.0 & 0.0 & 1.0 \\ \mathrm{Mn3:}& 0.0 & 0.0 & 1.0 \\ \mathrm{Mn4:}& 0.0 & 0.0 & 1.0 \end{array}\right)$&$Pn'm'a$& $m'm'm$&$\sigma_{xy}$\\
$A_{u}$& $M_u$ &$\left(\begin{array}{rrrr} \mathrm{Mn1:}& -0.410228 & 0.000000 & 0.911982 \\ \mathrm{Mn2:}& -0.410228 & 0.000000 & -0.911982 \\ \mathrm{Mn3:} & 0.410228 & 0.000000 & 0.911982 \\ \mathrm{Mn4:}&  0.410228 & 0.000000 & -0.911982 \end{array}\right)$& $Pn'm'a'$&$m'm'm'$&NONE\\   
$A_{u}$& $M_v$ &$\left(\begin{array}{rrrr} \mathrm{Mn1:} &0.911982 & 0.000000 & 0.410228 \\ \mathrm{Mn2:} & 0.911982 & 0.000000 & -0.410228 \\ \mathrm{Mn3:} &-0.911982 & 0.000000 & 0.410228 \\ \mathrm{Mn4:}& -0.911982 & 0.000000 & -0.410228 \end{array}\right)$& $Pn'm'a'$&$m'm'm'$&NONE\\   
$B_{2u}$& $M_{zx}$ &$\left(\begin{array}{rrrr} \mathrm{Mn1:}& -0.668816 & 0.000000 & -0.743428 \\ \mathrm{Mn2:} &0.668816 & 0.000000 & -0.743428 \\ \mathrm{Mn3:} &-0.668816 & 0.000000 & 0.743428 \\ \mathrm{Mn4:}& 0.668816 & 0.000000 & 0.743428 \end{array}\right)$&$Pnm'a$& $mmm'$&NONE\\   
$B_{3u}$& $T_x$ &$\left(\begin{array}{rrrr} \mathrm{Mn1:} &0.0 & -1.0 &0.0 \\ \mathrm{Mn2:} &0.0 & 1.0 & 0.0 \\ \mathrm{Mn3:}& 0.0 & -1.0 & 0.0 \\ \mathrm{Mn4:}& 0.0 & 1.0 & 0.0 \end{array}\right)$& $Pn'ma$&$mmm'$&NONE\\
$B_{2u}$& $T_y$ &$\left(\begin{array}{rrrr} \mathrm{Mn1:}& 0.743428 & 0.000000 & -0.668816 \\ \mathrm{Mn2:}& -0.743428 & 0.000000 & -0.668816 \\ \mathrm{Mn3:}& 0.743428 & 0.000000 & 0.668816 \\ \mathrm{Mn4:} &-0.743428 & 0.000000 & 0.668816 \end{array}\right)$& $Pnm'a$&$mmm'$&NONE\\
$B_{1u}$& $T_z$ &$\left(\begin{array}{rrrr} \mathrm{Mn1:}& 0.0 & 1.0 &0.0 \\ \mathrm{Mn2:}& 0.0 & 1.0 & 0.0 \\ \mathrm{Mn3:}& 0.0 & -1.0 & 0.0 \\ \mathrm{Mn4:}& 0.0 & -1.0 & 0.0 \end{array}\right)$& $Pnma'$&$mmm'$&NONE\\
$A_{g}$& $T_{u}$ &$\left(\begin{array}{rrrr} \mathrm{Mn1:}& 0.0 & 1.0 &0.0 \\  \mathrm{Mn2:} &0.0 & -1.0 & 0.0 \\ \mathrm{Mn3:} &0.0 & -1.0 & 0.0 \\ \mathrm{Mn4:}& 0.0 & 1.0 & 0.0 \end{array}\right)$& $Pnma$&$mmm$&NONE\\   
$B_{3g}$& $T_{yz}$ &$\left(\begin{array}{rrrr} \mathrm{Mn1:} &0.0 & 0.0 & 1.0 \\ \mathrm{Mn2:}& 0.0 & 0.0 & -1.0 \\ \mathrm{Mn3:}& 0.0 & 0.0 & -1.0 \\ \mathrm{Mn4:}& 0.0 & 0.0 & 1.0 \end{array}\right)$&$Pnm'a'$& $m'm'm$&$\sigma_{yz}$\\   
$B_{1g}$& $T_{xy}$ &$\left(\begin{array}{rrrr} \mathrm{Mn1:} &-1.0 & 0.0 & 0.0 \\ \mathrm{Mn2:} & 1.0 & 0.0 & 0.0 \\ \mathrm{Mn3:}& 1.0 & 0.0 & 0.0 \\ \mathrm{Mn4:} &-1.0 & 0.0 & 0.0 \end{array}\right)$& $Pn'm'a$&$m'm'm$&$\sigma_{xy}$\\   
\hline
\hline
\end{tabular}  
\end{table*}
\begin{figure} 
\centering 
\includegraphics[width=7.5cm]{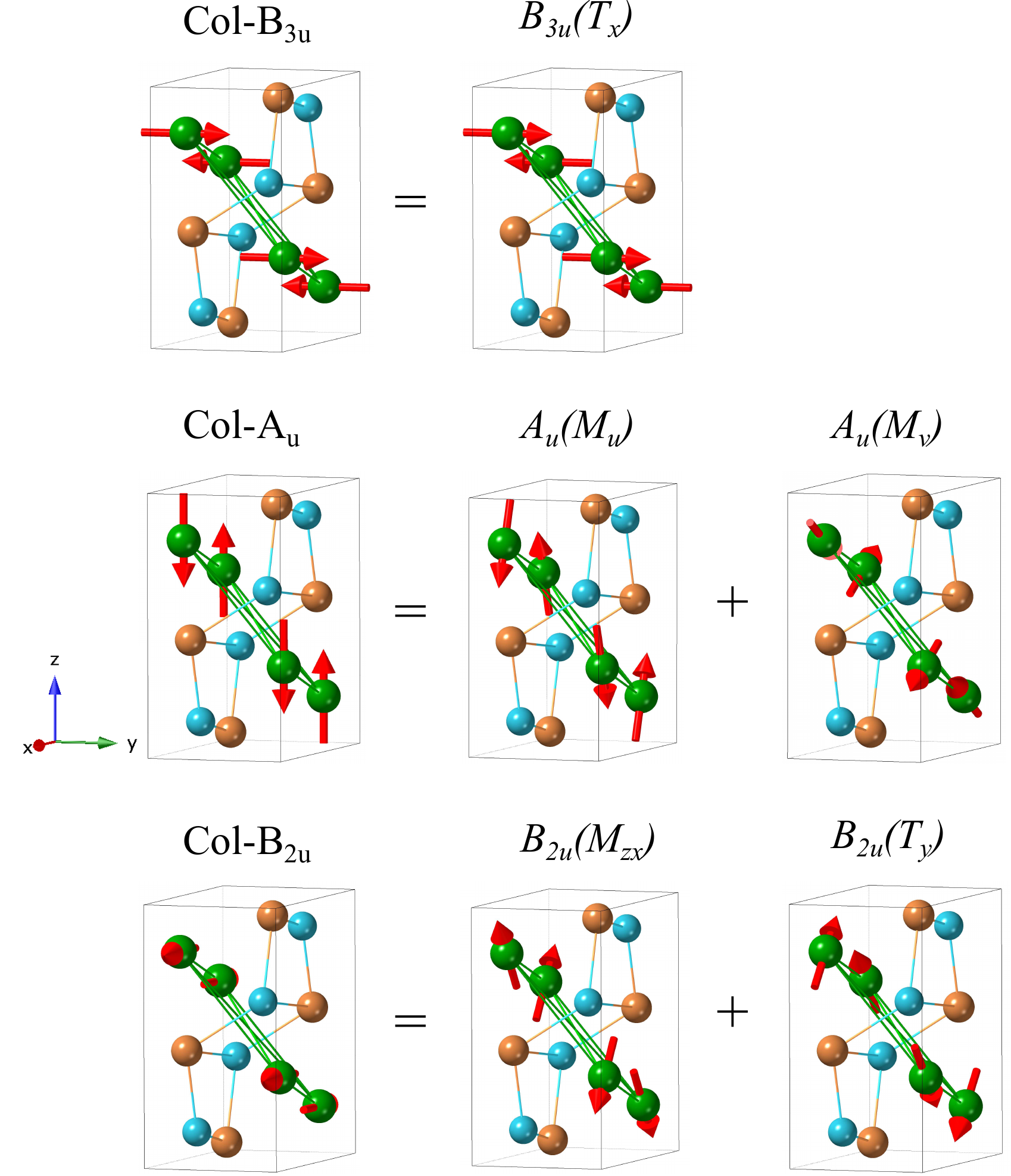}\\
\captionof{figure}{Linear combinations of the magnetic-structure bases to obtain the collinear-AFM states.}
\label{fig:oyzx}
\end{figure}
\begin{table}
\captionof{table}{Symmetry operators of magnetic point groups in the Col-B$_{3\mathrm{u}}$,  Col-A$_{\mathrm{u}}$, and Col-B$_{2\mathrm{u}}$ structures.}
\label{tab:operators}
\begin{tabular}{P{2.5cm}P{2.5cm}P{2.5cm}} 
\hline
\hline
 Col-B$_{3\mathrm{u}}$ &  Col-A$_{\mathrm{u}}$ & Col-B$_{2\mathrm{u}}$  \\
 \hline
 $E$  & $E$ & $E$ \\
$C_{2x}$& $C_{2x}$  & $C_{2y}$ \\
$PC_{2y}$ & $C_{2y}$ & $PC_{2x}$ \\
$PC_{2z}$ & $C_{2z}$ & $PC_{2z}$ \\
$TC_{2y}$  & $TP$ & $TC_{2x}$ \\
$TC_{2z}$  &$TPC_{2x}$& $TC_{2z}$ \\
$TP$  & $TPC_{2y}$  & $TP$ \\
$TPC_{2x}$  & $TPC_{2z}$ & $TPC_{2y}$\\
\hline
\hline
\end{tabular}  
\end{table}

Orthorhombic CuMnAs crystallizes in space group {\it Pnma} (No.~62), whose point group is $D_{2h}$, and Cu, Mn, and As are all located in $4c$-site.
Figure~\ref{fig:orth} shows the energetically inequivalent magnetic structures with the ordering vector $\bm{q}=\bm{0}$, classified according to the irreducible representations (IREPs) of the $D_{2h}$ point group, generated by the cluster multipole expansion scheme of Ref.~\onlinecite{cmpgeneration}. 
The generated magnetic structures are consistent with the decomposition of magnetic representation $D^{(\mathrm{mag})}$ of the $4c$-site in space group {\it Pnma} into IREPs as follows:
\begin{align}
 D^{(\mathrm{mag})} &=   A_{g} \oplus 2B_{1g} \oplus B_{2g} \oplus 2B_{3g} \nonumber \\
& \oplus 2A_{u} \oplus B_{1u} \oplus 2B_{2u} \oplus B_{3u}.
\end{align}

We list magnetic moments on four manganese atoms, the magnetic space group, magnetic point groups, and finite AHC tensor components in Table \ref{tab:orthonormal}.
The magnetic structures belonging to ungerade IREP with subscript ``$u$" forbid to have a finite AHC since they preserve $TP$ symmetry, which imposes the vanishing of Berry curvature at arbitrary $\bm{k}$-points in the entire BZ \cite{Martinez_2015,cmp2017}.

The magnetic toroidal (MT) dipole $B_{3u}(T_x)$ in Fig.~\ref{fig:orth}, corresponding to a collinear-AFM with the magnetic moment orientation along the $y$-axis, was experimentally observed \cite{cumnas6}. We refer this collinear-AFM structure as Col-B$_{3\mathrm{u}}$ structure.
The collinear-AFM obtained by the linear combination of magnetic (M) quadrupoles $M_u$ and $M_v$ belonging to $A_{u}$-IREP, referred as Col-A$_{\mathrm{u}}$, has local magnetic moments along $z$-axis.
Another collinear-AFM obtained by a linear combination of M quadrupole $M_{zx}$ and MT quadrupole $T_y$ belonging to $B_{2u}$-IREP, referred as Col-B$_{2\mathrm{u}}$, has local magnetic moments along $x$-axis.
Three magnetic structures Col-B$_{3\mathrm{u}}$, Col-A$_{\mathrm{u}}$, and Col-B$_{2\mathrm{u}}$ are illustrated in Fig.~\ref{fig:oyzx} and their symmetry operators of the magnetic point groups  are listed in Table \ref{tab:operators}.

To recognize finite SHC tensor components as well as to understand relations of spin Berry curvature with magnetic symmetry operations, we provide a detailed analysis about constraint on the spin Berry curvature in $k$-space for some representative symmetries.
Considering unitary symmetry operators $R$ and anti-unitary symmetry operators $A=TR$ with their representation matrices $D^j$ for charge current and $D^{j^s}$ for spin current, respectively, we have relations for the spin Berry curvature for the unitary operation as follows:
\begin{equation}
\label{equ:spinberryu}
\begin{split}
\Omega^{\mathrm{S},\gamma}_{n,\alpha \beta}(R\bm{k})= \sum_{\alpha' \beta' \gamma'}D^{j_s}_{\alpha \gamma \alpha' \gamma' }(R)D^{j}_{\beta \beta'}(R)\Omega^{\mathrm{S},\gamma'}_{n,\alpha' \beta'}(\bm{k})
\end{split}
\end{equation}
and for anti-unitary operation:%
\begin{equation}
\label{equ:spinberryau}
\begin{split}
\Omega^{\mathrm{S},\gamma}_{n,\alpha \beta}(A\bm{k})=-\sum_{\alpha ' \beta ' \gamma '}D^{j_s}_{\alpha \gamma \alpha' \gamma'}(A)D^{j}_{\beta \beta'}(A)\Omega^{\mathrm{S},\gamma'}_{n,\alpha'\beta'}(\bm{k})\ .
\end{split}
\end{equation}
We list the relations of the reciprocal coordinate $\bm{k}$ and spin Berry curvature under the unitary and antiunitary transformation for some symmetry operations of cubic, tetragonal, and orthorhombic structures in Table \ref{tab:spinBerry}.
\begin{table*}
\captionof{table}{ Relations of the reciprocal coordinate $\bm{k}$ and spin Berry curvature under the unitary and antiunitary transformation for some representative symmetries in the cubic, tetragonal, and orthorhombic systems. $C_{n\mu}, P, T$ indicate the $n$-fold rotation operator along the $\mu$ axis, the spatial inversion, and the time-reversal operator, respectively. For example, $\Omega^{\mathrm{S},y}_{yy}$ with $[\dagger]$ in the table means $\Omega^{\mathrm{S},y}_{yy}(\bm{k}')=\Omega^{\mathrm{S},x}_{xx}(\bm{k})$ under the $C_{4z}$ symmetry with $\bm{k}'=(-k_y,k_x,k_z)$.}
\label{tab:spinBerry}
a)For $\underline{{\bm \Omega }}^{\mathrm{S},x}$\\
\begin{tabular}{P{1.2cm}P{3.4cm}P{1.2cm}P{1.2cm}P{1.2cm}P{1.2cm}P{1.2cm}P{1.2cm}P{1.2cm}P{1.2cm}P{1.2cm}}
\hline
\hline
   & $\bm{k}'=R\bm{k}$ or $A\bm{k}$ & $\Omega^{\mathrm{S},x}_{xx}$ & $\Omega^{\mathrm{S},x}_{xy}$  &  $\Omega^{\mathrm{S},x}_{xz}$ & $\Omega^{\mathrm{S},x}_{yx}$    &$\Omega^{\mathrm{S},x}_{yy}$ &  $\Omega^{\mathrm{S},x}_{yz}$  & $\Omega^{\mathrm{S},x}_{zx}$    &$\Omega^{\mathrm{S},x}_{zy}$ &  $\Omega^{\mathrm{S},x}_{zz}$    \\
 \hline
$P$  & $(-k_x,-k_y,-k_z)$ & $\Omega^{\mathrm{S},x}_{xx}$ & $\Omega^{\mathrm{S},x}_{xy}$  &  $\Omega^{\mathrm{S},x}_{xz}$ & $\Omega^{\mathrm{S},x}_{yx}$    &$\Omega^{\mathrm{S},x}_{yy}$ &  $\Omega^{\mathrm{S},x}_{yz}$  & $\Omega^{\mathrm{S},x}_{zx}$    &$\Omega^{\mathrm{S},x}_{zy}$ &  $\Omega^{\mathrm{S},x}_{zz}$ \\
$T$  &  $(-k_x,-k_y,-k_z)$   & $\Omega^{\mathrm{S},x}_{xx}$ & $\Omega^{\mathrm{S},x}_{xy}$  &  $\Omega^{\mathrm{S},x}_{xz}$ & $\Omega^{\mathrm{S},x}_{yx}$    &$\Omega^{\mathrm{S},x}_{yy}$ &  $\Omega^{\mathrm{S},x}_{yz}$  & $\Omega^{\mathrm{S},x}_{zx}$    &$\Omega^{\mathrm{S},x}_{zy}$ &  $\Omega^{\mathrm{S},x}_{zz}$\\
$TP$  & $(k_x,k_y,k_z)$  & $\Omega^{\mathrm{S},x}_{xx}$ & $\Omega^{\mathrm{S},x}_{xy}$  &  $\Omega^{\mathrm{S},x}_{xz}$ & $\Omega^{\mathrm{S},x}_{yx}$    &$\Omega^{\mathrm{S},x}_{yy}$ &  $\Omega^{\mathrm{S},x}_{yz}$  & $\Omega^{\mathrm{S},x}_{zx}$    &$\Omega^{\mathrm{S},x}_{zy}$ &  $\Omega^{\mathrm{S},x}_{zz}$  \\
$C_{2x}$  & $(k_x,-k_y,-k_z)$   & $\Omega^{\mathrm{S},x}_{xx}$ & $-\Omega^{\mathrm{S},x}_{xy}$  &  $-\Omega^{\mathrm{S},x}_{xz}$ & $-\Omega^{\mathrm{S},x}_{yx}$    &$\Omega^{\mathrm{S},x}_{yy}$ &  $\Omega^{\mathrm{S},x}_{yz}$  & $-\Omega^{\mathrm{S},x}_{zx}$    &$\Omega^{\mathrm{S},x}_{zy}$ &  $\Omega^{\mathrm{S},x}_{zz}$ \\
$C_{2y}$  & $(-k_x,k_y,-k_z)$   & $-\Omega^{\mathrm{S},x}_{xx}$ & $-\Omega^{\mathrm{S},x}_{xy}$  &  $-\Omega^{\mathrm{S},x}_{xz}$ & $-\Omega^{\mathrm{S},x}_{yx}$    &$-\Omega^{\mathrm{S},x}_{yy}$ &  $\Omega^{\mathrm{S},x}_{yz}$  & $-\Omega^{\mathrm{S},x}_{zx}$    &$\Omega^{\mathrm{S},x}_{zy}$ &  $-\Omega^{\mathrm{S},x}_{zz}$ \\
$C_{2z}$  & $(-k_x,-k_y,k_z)$   & $-\Omega^{\mathrm{S},x}_{xx}$ & $-\Omega^{\mathrm{S},x}_{xy}$  &  $-\Omega^{\mathrm{S},x}_{xz}$ & $-\Omega^{\mathrm{S},x}_{yx}$    &$-\Omega^{\mathrm{S},x}_{yy}$ &  $\Omega^{\mathrm{S},x}_{yz}$  & $-\Omega^{\mathrm{S},x}_{zx}$    &$\Omega^{\mathrm{S},x}_{zy}$ &  $-\Omega^{\mathrm{S},x}_{zz}$ \\
$C_{2[1\bar{1}0]}$          & $(-k_y,-k_x,-k_z)$   & $-\Omega^{\mathrm{S},y}_{yy}$ & $-\Omega^{\mathrm{S},y}_{yx}$  &  $-\Omega^{\mathrm{S},y}_{yz}$ & $-\Omega^{\mathrm{S},y}_{xy}$    &$-\Omega^{\mathrm{S},y}_{xx}$ &  $-\Omega^{\mathrm{S},y}_{xz}$  & $-\Omega^{\mathrm{S},y}_{zy}$    &$-\Omega^{\mathrm{S},y}_{zx}$ &  $-\Omega^{\mathrm{S},y}_{zz}$\\
$C_{3[111]}$           & $(k_z,k_x,k_y)$   & $\Omega^{\mathrm{S},z}_{zz}$ & $\Omega^{\mathrm{S},z}_{zx}$  &  $\Omega^{\mathrm{S},z}_{zy}$ & $\Omega^{\mathrm{S},z}_{xz}$    &$\Omega^{\mathrm{S},z}_{xx}$ &  $\Omega^{\mathrm{S},z}_{xy}$  & $\Omega^{\mathrm{S},z}_{yz}$    &$\Omega^{\mathrm{S},z}_{yx}$ &  $\Omega^{\mathrm{S},z}_{yy}$\\
$C_{4z}$               & $(-k_y,k_x,k_z)$  & $\Omega^{\mathrm{S},y}_{yy}$$[\dagger]$ & $-\Omega^{\mathrm{S},y}_{yx}$  &  $\Omega^{\mathrm{S},y}_{yz}$ & $-\Omega^{\mathrm{S},y}_{xy}$    &$\Omega^{\mathrm{S},y}_{xx}$ &  $-\Omega^{\mathrm{S},y}_{xz}$  & $\Omega^{\mathrm{S},y}_{zy}$    &$-\Omega^{\mathrm{S},y}_{zx}$ &  $\Omega^{\mathrm{S},y}_{zz}$\\
\hline
\hline
\end{tabular}\\  
b)For $\underline{{\bm \Omega }}^{\mathrm{S},y}$\\
\begin{tabular}{P{1.2cm}P{3.4cm}P{1.2cm}P{1.2cm}P{1.2cm}P{1.2cm}P{1.2cm}P{1.2cm}P{1.2cm}P{1.2cm}P{1.2cm}}
\hline
\hline
   & $\bm{k}'=R\bm{k}$ or $A\bm{k}$ & $\Omega^{\mathrm{S},y}_{xx}$ & $\Omega^{\mathrm{S},y}_{xy}$  &  $\Omega^{\mathrm{S},y}_{xz}$ & $\Omega^{\mathrm{S},y}_{yx}$    &$\Omega^{\mathrm{S},y}_{yy}$ &  $\Omega^{\mathrm{S},y}_{yz}$  & $\Omega^{\mathrm{S},y}_{zx}$    &$\Omega^{\mathrm{S},y}_{zy}$ &  $\Omega^{\mathrm{S},y}_{zz}$    \\
 \hline
$P$  & $(-k_x,-k_y,-k_z)$ & $\Omega^{\mathrm{S},y}_{xx}$ & $\Omega^{\mathrm{S},y}_{xy}$  &  $\Omega^{\mathrm{S},y}_{xz}$ & $\Omega^{\mathrm{S},y}_{yx}$    &$\Omega^{\mathrm{S},y}_{yy}$ &  $\Omega^{\mathrm{S},y}_{yz}$  & $\Omega^{\mathrm{S},y}_{zx}$    &$\Omega^{\mathrm{S},y}_{zy}$ &  $\Omega^{\mathrm{S},y}_{zz}$ \\
$T$  &  $(-k_x,-k_y,-k_z)$   & $\Omega^{\mathrm{S},y}_{xx}$ & $\Omega^{\mathrm{S},y}_{xy}$  &  $\Omega^{\mathrm{S},y}_{xz}$ & $\Omega^{\mathrm{S},y}_{yx}$    &$\Omega^{\mathrm{S},y}_{yy}$ &  $\Omega^{\mathrm{S},y}_{yz}$  & $\Omega^{\mathrm{S},y}_{zx}$    &$\Omega^{\mathrm{S},y}_{zy}$ &  $\Omega^{\mathrm{S},y}_{zz}$\\
$TP$  & $(k_x,k_y,k_z)$  & $\Omega^{\mathrm{S},y}_{xx}$ & $\Omega^{\mathrm{S},y}_{xy}$  &  $\Omega^{\mathrm{S},y}_{xz}$ & $\Omega^{\mathrm{S},y}_{yx}$    &$\Omega^{\mathrm{S},y}_{yy}$ &  $\Omega^{\mathrm{S},y}_{yz}$  & $\Omega^{\mathrm{S},y}_{zx}$    &$\Omega^{\mathrm{S},y}_{zy}$ &  $\Omega^{\mathrm{S},y}_{zz}$  \\
$C_{2x}$  & $(k_x,-k_y,-k_z)$   & $-\Omega^{\mathrm{S},y}_{xx}$ & $-\Omega^{\mathrm{S},y}_{xy}$  &  $\Omega^{\mathrm{S},y}_{xz}$ & $-\Omega^{\mathrm{S},y}_{yx}$    &$-\Omega^{\mathrm{S},y}_{yy}$ &  $-\Omega^{\mathrm{S},y}_{yz}$  & $\Omega^{\mathrm{S},y}_{zx}$    &$-\Omega^{\mathrm{S},y}_{zy}$ &  $-\Omega^{\mathrm{S},y}_{zz}$ \\
$C_{2y}$  & $(-k_x,k_y,-k_z)$   & $\Omega^{\mathrm{S},y}_{xx}$ & $-\Omega^{\mathrm{S},y}_{xy}$  &  $\Omega^{\mathrm{S},y}_{xz}$ & $-\Omega^{\mathrm{S},y}_{yx}$    &$\Omega^{\mathrm{S},y}_{yy}$ &  $-\Omega^{\mathrm{S},y}_{yz}$  & $\Omega^{\mathrm{S},y}_{zx}$    &$-\Omega^{\mathrm{S},y}_{zy}$ &  $\Omega^{\mathrm{S},y}_{zz}$ \\
$C_{2z}$  & $(-k_x,-k_y,k_z)$   & $-\Omega^{\mathrm{S},y}_{xx}$ & $-\Omega^{\mathrm{S},y}_{xy}$  &  $\Omega^{\mathrm{S},y}_{xz}$ & $-\Omega^{\mathrm{S},y}_{yx}$    &$-\Omega^{\mathrm{S},y}_{yy}$ &  $-\Omega^{\mathrm{S},y}_{yz}$  & $\Omega^{\mathrm{S},y}_{zx}$    &$-\Omega^{\mathrm{S},y}_{zy}$ &  $-\Omega^{\mathrm{S},y}_{zz}$ \\
$C_{2[1\bar{1}0]}$          & $(-k_y,-k_x,-k_z)$   & $-\Omega^{\mathrm{S},x}_{yy}$ & $-\Omega^{\mathrm{S},x}_{yx}$  &  $-\Omega^{\mathrm{S},x}_{yz}$ & $-\Omega^{\mathrm{S},x}_{xy}$    &$-\Omega^{\mathrm{S},x}_{xx}$ &  $-\Omega^{\mathrm{S},x}_{xz}$  & $-\Omega^{\mathrm{S},x}_{zy}$    &$-\Omega^{\mathrm{S},x}_{zy}$ &  $-\Omega^{\mathrm{S},x}_{zz}$\\
$C_{3[111]}$           & $(k_z,k_x,k_y)$   & $\Omega^{\mathrm{S},x}_{zz}$ & $\Omega^{\mathrm{S},x}_{zx}$  &  $\Omega^{\mathrm{S},x}_{zy}$ & $\Omega^{\mathrm{S},x}_{xz}$    &$\Omega^{\mathrm{S},x}_{xx}$ &  $\Omega^{\mathrm{S},x}_{xy}$  & $\Omega^{\mathrm{S},x}_{yz}$    &$\Omega^{\mathrm{S},x}_{yx}$ &  $\Omega^{\mathrm{S},x}_{yy}$\\
$C_{4z}$               & $(-k_y,k_x,k_z)$  & -$\Omega^{\mathrm{S},x}_{yy}$ & $\Omega^{\mathrm{S},x}_{yx}$  &  $-\Omega^{\mathrm{S},x}_{yz}$ & $\Omega^{\mathrm{S},x}_{xy}$    &$-\Omega^{\mathrm{S},x}_{xx}$ &  $\Omega^{\mathrm{S},x}_{xz}$  & $-\Omega^{\mathrm{S},x}_{zy}$    &$\Omega^{\mathrm{S},x}_{zx}$ &  $-\Omega^{\mathrm{S},x}_{zz}$\\
\hline
\hline
\end{tabular}  \\
c)For $\underline{{\bm \Omega }}^{\mathrm{S},z}$\\
\begin{tabular}{P{1.2cm}P{3.4cm}P{1.2cm}P{1.2cm}P{1.2cm}P{1.2cm}P{1.2cm}P{1.2cm}P{1.2cm}P{1.2cm}P{1.2cm}}
\hline
\hline
   & $\bm{k}'=R\bm{k}$ or $A\bm{k}$ & $\Omega^{\mathrm{S},z}_{xx}$ & $\Omega^{\mathrm{S},z}_{xy}$  &  $\Omega^{\mathrm{S},z}_{xz}$ & $\Omega^{\mathrm{S},z}_{yx}$    &$\Omega^{\mathrm{S},z}_{yy}$ &  $\Omega^{\mathrm{S},z}_{yz}$  & $\Omega^{\mathrm{S},z}_{zx}$    &$\Omega^{\mathrm{S},z}_{zy}$ &  $\Omega^{\mathrm{S},z}_{zz}$    \\
 \hline
$P$  & $(-k_x,-k_y,-k_z)$ & $\Omega^{\mathrm{S},z}_{xx}$ & $\Omega^{\mathrm{S},z}_{xy}$  &  $\Omega^{\mathrm{S},z}_{xz}$ & $\Omega^{\mathrm{S},z}_{yx}$    &$\Omega^{\mathrm{S},z}_{yy}$ &  $\Omega^{\mathrm{S},z}_{yz}$  & $\Omega^{\mathrm{S},z}_{zx}$    &$\Omega^{\mathrm{S},z}_{zy}$ &  $\Omega^{\mathrm{S},z}_{zz}$ \\
$T$  &  $(-k_x,-k_y,-k_z)$   & $\Omega^{\mathrm{S},z}_{xx}$ & $\Omega^{\mathrm{S},z}_{xy}$  &  $\Omega^{\mathrm{S},z}_{xz}$ & $\Omega^{\mathrm{S},z}_{yx}$    &$\Omega^{\mathrm{S},z}_{yy}$ &  $\Omega^{\mathrm{S},z}_{yz}$  & $\Omega^{\mathrm{S},z}_{zx}$    &$\Omega^{\mathrm{S},z}_{zy}$ &  $\Omega^{\mathrm{S},z}_{zz}$\\
$TP$  & $(k_x,k_y,k_z)$  & $\Omega^{\mathrm{S},z}_{xx}$ & $\Omega^{\mathrm{S},z}_{xy}$  &  $\Omega^{\mathrm{S},z}_{xz}$ & $\Omega^{\mathrm{S},z}_{yx}$    &$\Omega^{\mathrm{S},z}_{yy}$ &  $\Omega^{\mathrm{S},z}_{yz}$  & $\Omega^{\mathrm{S},z}_{zx}$    &$\Omega^{\mathrm{S},z}_{zy}$ &  $\Omega^{\mathrm{S},z}_{zz}$  \\
$C_{2x}$  & $(k_x,-k_y,-k_z)$   & $-\Omega^{\mathrm{S},z}_{xx}$ & $\Omega^{\mathrm{S},z}_{xy}$  &  $-\Omega^{\mathrm{S},z}_{xz}$ & $\Omega^{\mathrm{S},z}_{yx}$    &$-\Omega^{\mathrm{S},z}_{yy}$ &  $-\Omega^{\mathrm{S},z}_{yz}$  & $-\Omega^{\mathrm{S},z}_{zx}$    &$-\Omega^{\mathrm{S},z}_{zy}$ &  $-\Omega^{\mathrm{S},z}_{zz}$ \\
$C_{2y}$  & $(-k_x,k_y,-k_z)$   & $-\Omega^{\mathrm{S},z}_{xx}$ & $\Omega^{\mathrm{S},z}_{xy}$  &  $-\Omega^{\mathrm{S},z}_{xz}$ & $\Omega^{\mathrm{S},z}_{yx}$    &$-\Omega^{\mathrm{S},z}_{yy}$ &  $-\Omega^{\mathrm{S},z}_{yz}$  & $-\Omega^{\mathrm{S},z}_{zx}$    &$-\Omega^{\mathrm{S},z}_{zy}$ &  $-\Omega^{\mathrm{S},z}_{zz}$ \\
$C_{2z}$  & $(-k_x,-k_y,k_z)$   & $\Omega^{\mathrm{S},z}_{xx}$ & $\Omega^{\mathrm{S},z}_{xy}$  &  $-\Omega^{\mathrm{S},z}_{xz}$ & $\Omega^{\mathrm{S},z}_{yx}$    &$\Omega^{\mathrm{S},z}_{yy}$ &  $-\Omega^{\mathrm{S},z}_{yz}$  & $-\Omega^{\mathrm{S},z}_{zx}$    &$-\Omega^{\mathrm{S},z}_{zy}$ &  $\Omega^{\mathrm{S},z}_{zz}$ \\
$C_{2[1\bar{1}0]}$          & $(-k_y,-k_x,-k_z)$   & $-\Omega^{\mathrm{S},z}_{yy}$ & $-\Omega^{\mathrm{S},z}_{yx}$  &  $-\Omega^{\mathrm{S},z}_{yz}$ & $-\Omega^{\mathrm{S},z}_{xy}$    &$-\Omega^{\mathrm{S},z}_{xx}$ &  $-\Omega^{\mathrm{S},z}_{xz}$  & $-\Omega^{\mathrm{S},z}_{zy}$    &$-\Omega^{\mathrm{S},z}_{zx}$ &  $-\Omega^{\mathrm{S},z}_{zz}$\\
$C_{3[111]}$           & $(k_z,k_x,k_y)$   & $\Omega^{\mathrm{S},y}_{zz}$ & $\Omega^{\mathrm{S},y}_{zx}$  &  $\Omega^{\mathrm{S},y}_{zy}$ & $\Omega^{\mathrm{S},y}_{xz}$    &$\Omega^{\mathrm{S},y}_{xx}$ &  $\Omega^{\mathrm{S},y}_{xy}$  & $\Omega^{\mathrm{S},y}_{yz}$    &$\Omega^{\mathrm{S},y}_{yx}$ &  $\Omega^{\mathrm{S},y}_{yy}$\\
$C_{4z}$               & $(-k_y,k_x,k_z)$  & $\Omega^{\mathrm{S},z}_{yy}$ & $-\Omega^{\mathrm{S},z}_{yx}$  &  $\Omega^{\mathrm{S},z}_{yz}$ & $-\Omega^{\mathrm{S},z}_{xy}$    &$\Omega^{\mathrm{S},z}_{xx}$ &  $-\Omega^{\mathrm{S},z}_{xz}$  & $\Omega^{\mathrm{S},z}_{zy}$    &$-\Omega^{\mathrm{S},z}_{zx}$ &  $\Omega^{\mathrm{S},z}_{zz}$\\
\hline
\hline
\end{tabular}  
\end{table*}

As seen in the table, the spin Berry curvature keeps its sign even under the spatial inversion $P$, the time-reversal $T$, and $TP$ symmetry. As a result, SHC components can be finite after BZ integration in Eq.~(\ref{equ:shc}) even with these symmetries as for non-magnetic state.
In addition, spin Berry curvature cancels out after BZ integration for most of the  components under $C_{2x}$, $C_{2y}$, $C_{2z}$, or conjunction of these symmetry operators with $T$, $P$, or $TP$, leading to vanishing of several components of SHC tensor.
For example, we can derive following relations under $C_{2x}$, $PC_{2x}$, $TC_{2x}$, or $TPC_{2x}$ symmetries: 
\begin{equation}
\label{equ:relationsb}
\begin{split}
\sigma^{\mathrm{S},x}_{xy}&=\sigma^{\mathrm{S},x}_{yx}=\sigma^{\mathrm{S},x}_{xz}=\sigma^{\mathrm{S},x}_{zx}=0,\\
\sigma^{\mathrm{S},y}_{xx}&=\sigma^{\mathrm{S},y}_{yy}=\sigma^{\mathrm{S},y}_{zz}=\sigma^{\mathrm{S},y}_{yz}=\sigma^{\mathrm{S},y}_{zy}=0,\\
\sigma^{\mathrm{S},z}_{xx}&=\sigma^{\mathrm{S},z}_{yy}=\sigma^{\mathrm{S},z}_{zz}=\sigma^{\mathrm{S},z}_{yz}=\sigma^{\mathrm{S},z}_{zy}=0.\\
\end{split}
\end{equation}
Further finite SHC components can be seen from transformation relations of spin Berry curvature for other symmetry operations listed in Table \ref{tab:spinBerry}.
As the result from above symmetry analysis for symmetry operators in Table \ref{tab:operators}, we obtain SHC tensor components finite for the Col-B$_{3\mathrm{u}}$, Col-A$_{\mathrm{u}}$, and Col-B$_{2\mathrm{u}}$ magnetic structures in Table \ref{tab:shc}, which is consistent to the results shown in Ref.~\onlinecite{seemann}.
\begin{table}
\captionof{table}{The symmetry-imposed shapes of the SHC tensors for the Col-B$_{3\mathrm{u}}$, Col-A$_{\mathrm{u}}$, and Col-B$_{2\mathrm{u}}$ structures.}
\label{tab:shc}
\begin{tabular}{P{2.5cm}P{2.5cm}P{2.5cm}} 
\hline
\hline
 $\underline{{\bm \sigma }}^{\mathrm{S},x}$ &  $\underline{{\bm \sigma }}^{\mathrm{S},y}$ &  $\underline{{\bm \sigma }}^{\mathrm{S},z}$  \\
 \hline
  $ \left( \begin{array}{ccc} 0 & 0&0 \\ 0 & 0 & \sigma^{\mathrm{S},x}_{yz} \\  0 & \sigma^{\mathrm{S},x}_{zy} & 0 \end{array}\right)$ &   $ \left( \begin{array}{ccc} 0 & 0&\sigma^{\mathrm{S},y}_{xz} \\ 0 & 0 &0 \\ \sigma^{\mathrm{S},y}_{zx} & 0 & 0 \end{array}\right)$ & $ \left( \begin{array}{ccc} 0 & \sigma^{\mathrm{S},z}_{xy}&0 \\ \sigma^{\mathrm{S},z}_{yx} & 0 &0 \\ 0 & 0 & 0 \end{array}\right)$\\
\hline
\hline
\end{tabular}  
\end{table}
\section{Results}\label{results}
\subsection{Magnetic stability and ground state properties}
\begin{figure} 
\centering 
\includegraphics[width=7.5cm]{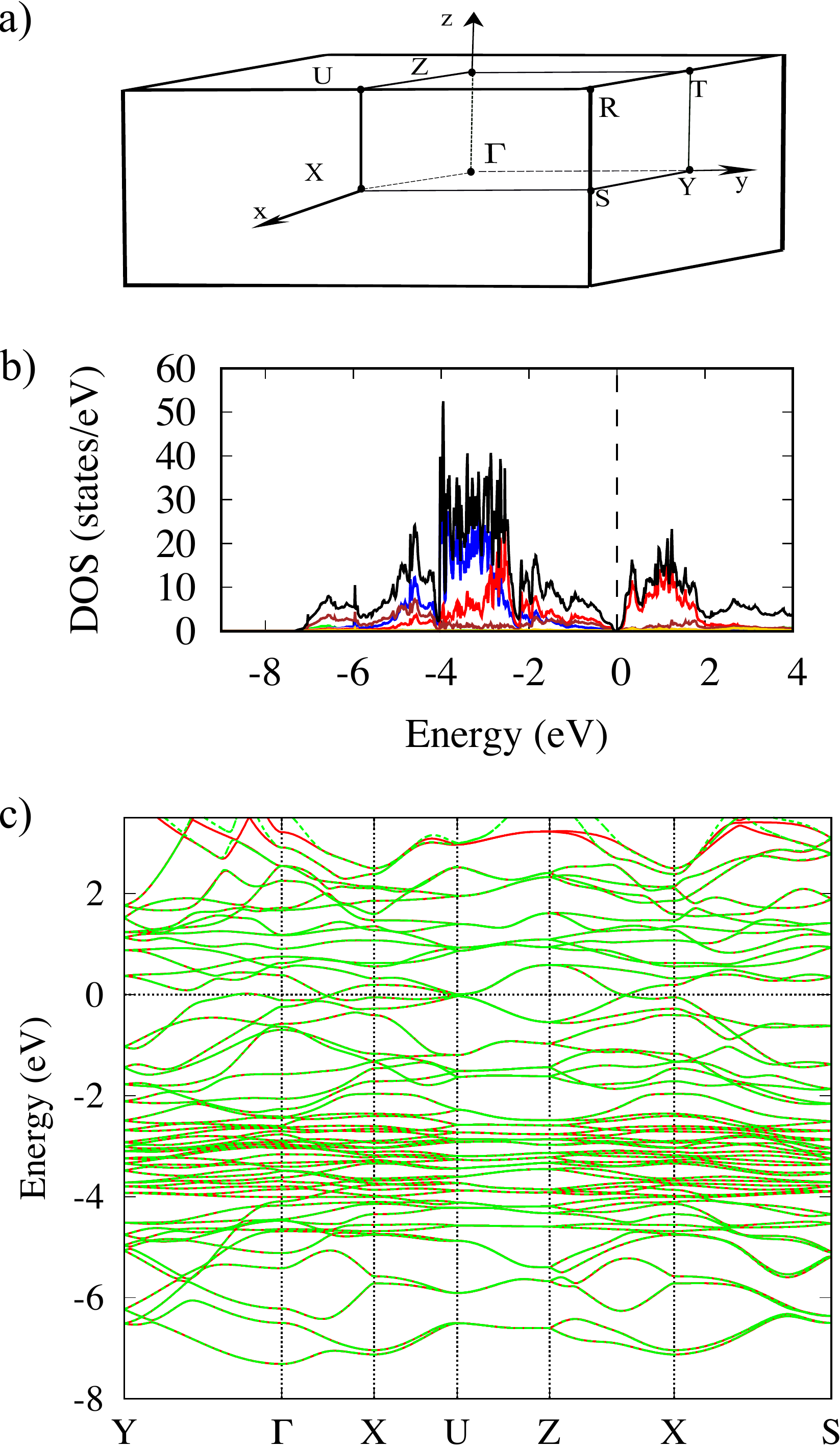}\\
\captionof{figure}{(a) The first BZ of a simple orthorhombic lattice with high-symmetry points. (b) The projected density of state of Col-B$_{3\mathrm{u}}$ structure for the orbitals Cu-4$s$ (green), Cu-$3d$ (blue), Mn-$4s$ (pink), Mn-$3d$ (red), As-$4p$ (brown), As-$4s$ (gold), and total density of state (black). (c) Energy bands from the first-principles calculations (red) and from Wannier interpolation (green) of Col-B$_{3\mathrm{u}}$ magnetic structure along high-symmetry lines.}
\label{fig:bandfitting}
\end{figure}
\begin{table}
\captionof{table}{Total magnetization $|M\mathrm{_{total}|(\mu_B)}$ of magnetic structures in Fig.~\ref{fig:orth} and Fig.~\ref{fig:oyzx} and the difference of total energy between each magnetic orderings and non-magnetic state $\mathrm{\Delta}E_{nm}$ (meV/u.c.) are listed. To easy discern, the different energies $\mathrm{\Delta}E$ (meV/u.c.) between each magnetic orderings with the most stable magnetic structure Col-B$_{2\mathrm{u}}$ are also shown in the last column.}
\label{tab:energy}
\begin{tabular}{P{1.9cm}P{1.9cm}P{2.0cm}P{2.0cm}} 
\hline
\hline
Structure &$|M\mathrm{_{total}|}$$(\mu_{\mathrm{B}})$ & $\mathrm{\Delta}E_{nm}$ (meV/u.c.) & $\mathrm{\Delta}E$ (meV/u.c.) \\
 \hline
 $B_{3g} (M_x)$ & $16.3$ & $-3878.5$ & $912.4$ \\
$B_{2g} (M_y)$ & $16.3$ & $-3878.3$ & $912.6$ \\
 $B_{1g} (M_z)$ &$16.3$ & $-3878.2$ & $912.7$ \\
$A_{u} (M_u) $ &$0.0$& $-4784.3$ & $6.6$ \\
 $A_{u} (M_v) $&$0.0$& $-4444.6$ & $346.3$ \\ 
$B_{2u} (M_{zx})$ &$0.0$& $-4771.5$ & $19.4$ \\
 Col-B$_{2\mathrm{u}}$ &$0.0$& $-4790.9$ & $0.0$ \\
 Col-B$_{3\mathrm{u}}$  &$0.0$& $-4790.7$ & $0.2$ \\
 Col-A$_{\mathrm{u}}$  &$0.0$& $-4790.5$ & $0.4$ \\
 $B_{2u} (T_y)$ &$0.0$& $-4784.9$ & $6.0$ \\
$B_{1u} (T_z)$ &$0.0$& $-4516.7$ & $272.2$ \\
$A_{g} (T_{u})$ &$0.0$& $-4750.9$ & $40.0$ \\
 $B_{3g} (T_{yz})$ &$0.0$& $-4750.7$ & $40.2$ \\
$B_{1g} (T_{xy})$ &$0.0$& $-4750.8$ & $40.1$ \\
\hline
\hline
\end{tabular}  
\end{table}

\begin{table}
\captionof{table}{Relaxed atomic positions for Col-B$_{3u}$ magnetic structure in CuMnAs.}
\label{tab:positions}
\begin{tabular}{P{2cm}P{2cm}P{2cm}P{2cm}} 
\hline
\hline
 Atoms &  $x$ & $y$ & $z$  \\
 \hline
Cu & 0.379&0.25 & 0.055 \\
Cu & 0.621&0.75&0.945 \\
Cu & 0.121&0.75&0.555 \\
Cu & 0.879&0.25&0.445 \\
Mn1 & 0.457 &0.25& 0.674 \\
Mn2 & 0.543 &0.75& 0.327 \\
Mn3 & 0.043 &0.75& 0.174 \\
Mn4 & 0.957 &0.25& 0.827 \\
As & 0.761 &0.25& 0.127\\
As & 0.239 &0.75& 0.873 \\
As & 0.739 &0.75& 0.627 \\
As &0.261 &0.25& 0.373 \\
\hline
\hline
\end{tabular}  
\end{table}

We calculate total energies for the magnetic configurations in Fig.~\ref{fig:orth} and Fig.~\ref{fig:oyzx} in the presence of SOC to evaluate stability of the magnetic structures as the way suggested recently~\cite{Huebsch}, setting a fully polarized valence state as the initial setting.
The obtained total energy differences and total magnetization are listed in Table \ref{tab:energy}.
The magnetic structures different only from magnetic anisotropy produce the energy differences only with considering SOC and have the same total energies if the first-principles calculations are 
implemented without SOC. 
As a result, Col-B$_{2\mathrm{u}}$, Col-B$_{3\mathrm{u}}$, and Col-A$_{\mathrm{u}}$ structures have close 
total energies as well as those for the FM structures $M_x$, $M_y$, and $M_z$ in Fig.~\ref{fig:orth} due to small SOC in Mn-3$d$ orbitals, as shown in Table \ref{tab:energy}.

The result in Table \ref{tab:energy} shows that Col-B$_{2\mathrm{u}}$, Col-B$_{3\mathrm{u}}$, and Col-A$_{\mathrm{u}}$ are stable structures with the energy difference within 0.4 meV and much lower than other magnetic structures. 
The small discrepancy of the most stable magnetic structure from experimentally observed one, that is the calculation predicting the most stable magnetic structure as Col-B$_{2\mathrm{u}}$ against the Col-B$_{3\mathrm{u}}$ reported in experiment \cite{cumnas6}, can be addressed to the low accuracy for magnetic anisotropy within GGA calculations~\cite{accuracy_gga}.
Hereinafter, we will focus on characteristics of Col-B$_{3\mathrm{u}}$, Col-A$_{\mathrm{u}}$, and Col-B$_{2\mathrm{u}}$ magnetic structures.
The magnetic moments after getting convergence for the calculations using the initial magnetic structure of Col-A$_{\mathrm{u}}$ and Col-B$_{2\mathrm{u}}$ have the magnetic moments slightly canted along 
$x$- and $z$- directions, respectively, while those for Col-B$_{3\mathrm{u}}$ are purely on the $y$-direction.
The difference in canting structure of magnetic alignments for different collinear AFM structures is understood from the geometrical degree of freedom of the magnetic alignment adopted in each IREP as illustrated in Fig.~\ref{fig:oyzx}.
 The atomic positions obtained for Col-B$_{3\mathrm{u}}$ structure is listed in Table \ref{tab:positions}, and differences of those for Col-A$_{\mathrm{u}}$ and Col-B$_{2\mathrm{u}}$ structures are within 0.005 \AA.

The BZ of orthorhombic CuMnAs is illustrated with high-symmetry points in Fig.~\ref{fig:bandfitting}(a).
Figure \ref{fig:bandfitting}(b) shows the projected density of state for the Col-B$_{\mathrm{3u}}$ structure as those for Col-A$_{\mathrm{u}}$ and Col-B$_{2\mathrm{u}}$ structures are similar with the plotted energy scale. All three collinear-AFM structures are semi-metallic and the Mn-3$d$ and Cu-3$d$ orbitals give large density near the Fermi level, implying their dominant role to the transport properties. 
The energy bands of tight-binding models reproduce the energy bands obtained by first-principles calculations within the energy interval from the lowest energy of the valence bands to 2 eV above the Fermi energy, as shown in Fig.~\ref{fig:bandfitting}(c).
\begin{figure} 
\centering 
\includegraphics[width=7.5cm]{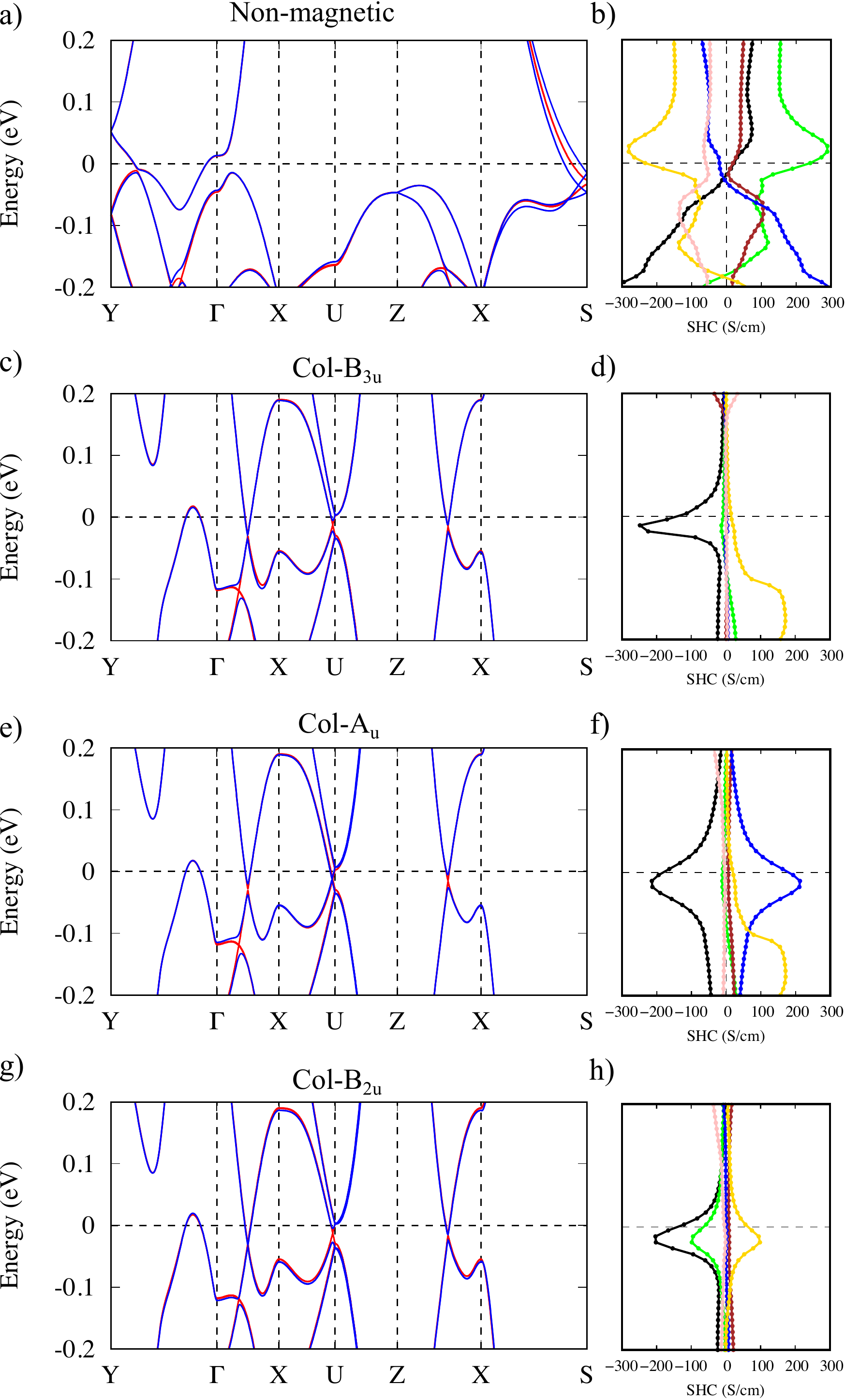}\\
\captionof{figure}{(a), (c), (e), and (g): Electronic band structures along high-symmetry lines without SOC (red) and with SOC (blue) in (a) nonmagnetic, (c) Col-B$_{3\mathrm{u}}$, (e) Col-A$_{\mathrm{u}}$, and (g) Col-B$_{2\mathrm{u}}$ states, respectively. (b), (d), (f), and (h): The SHC components $\sigma^{\mathrm{S},z}_{xy}$ (black),  $\sigma^{\mathrm{S},z}_{yx}$ (blue),$\sigma^{\mathrm{S},x}_{yz}$ (green), $\sigma^{\mathrm{S},x}_{zy}$ (gold), $\sigma^{\mathrm{S},y}_{zx}$ (brown), and $\sigma^{\mathrm{S},y}_{xz}$ (pink) in (b) non-magnetic, (d) Col-B$_{3\mathrm{u}}$, (f) Col-A$_{\mathrm{u}}$, and (h) Col-B$_{2\mathrm{u}}$ structures, respectively.}
\label{fig:bandshe}
\end{figure}

Figure \ref{fig:bandshe}(a), (c), (e), and (g) show the electronic band structures near the Fermi energy in the non-magnetic, 
Col-B$_{3\mathrm{u}}$ magnetic, Col-A$_{\mathrm{u}}$ magnetic, and Col-B$_{2\mathrm{u}}$ magnetic structures plotted both with and without SOC. 
These collinear AFM states are characterized with, at least, doubly degenerate electronic bands at all $k$ points due to preservation of $TP$ symmetry. As a result, crossing points in the energy bands are Dirac points, which are characterized with four-fold band degeneracy.
For the calculations neglecting SOC, these collinear-AFM structures have Dirac nodal lines in the (010) plane containing $\Gamma$ point as illustrated in Fig.~\ref{fig:kslice}(a) with small dispersion in the energy from 0 to about 30 meV. The nodal line degeneracy for Col-B$_{3\mathrm{u}}$ and Col-B$_{2\mathrm{u}}$ magnetic structures split entirely with considering the effect of SOC. Meanwhile, Col-A$_{\mathrm{u}}$ magnetic structure forms one pair of Dirac points, protected by screw rotation symmetry $S_{2z}$, on the nodal line even after considering the SOC, as investigated in Ref.~\cite{cumnas3}.
The two Dirac points of Col-A$_{\mathrm{u}}$ magnetic structure are located between X and U points with the coordinates D$_1$(0.50, 0.0, 0.47) and D$_2$(-0.50, 0.0, -0.47) at -12.7 meV from the Fermi level. 
%
\subsection{Topological degeneracy and spin Hall effect}
\begin{figure} 
\centering 
\includegraphics[width=7.5cm]{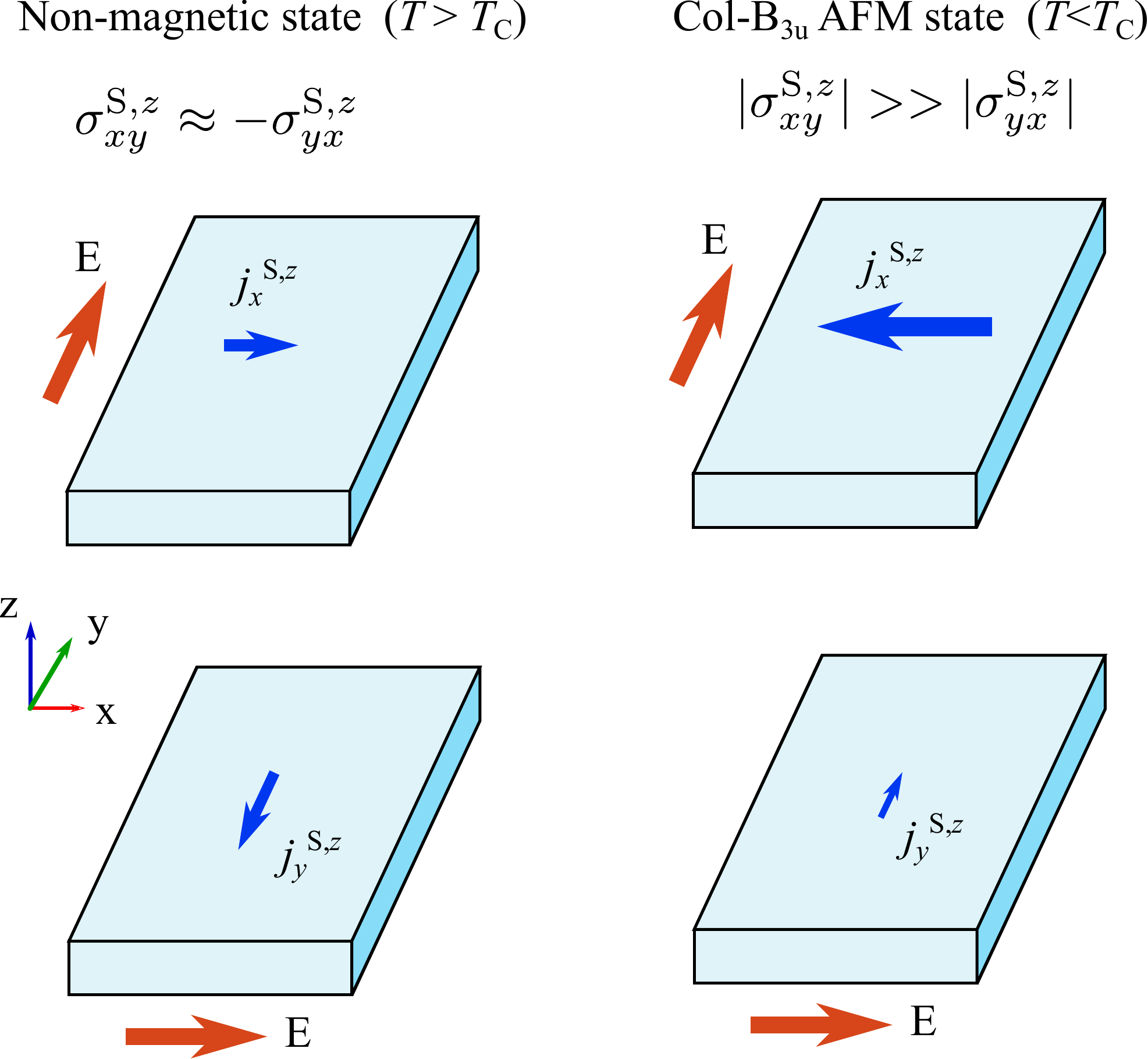}\\
\captionof{figure}{Illustration for the anisotropic SHE, which can be exploited for spintronic applications. Here, $E$ is the electric field, $j^{\mathrm{S},z}$ indicates the spin current with $z$-spin component, and $T_\mathrm{C}$ is the transition temperature from the AFM state to the non-magnetic state.}
\label{fig:illustrate}
\end{figure}
\begin{figure} 
\centering 
\includegraphics[width=7.0cm]{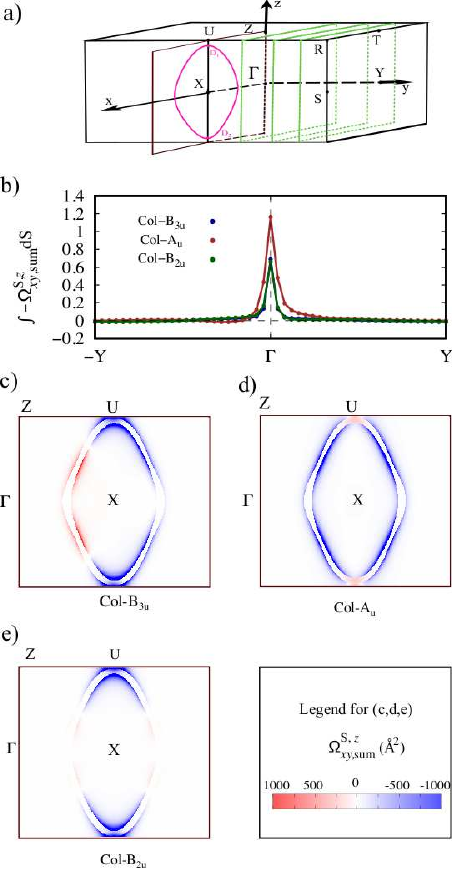}\\
\captionof{figure}{(a) Nodal line (pink) in (010) plane including $\Gamma$ in the Col-B$_{3\mathrm{u}}$, Col-A$_{\mathrm{u}}$, and Col-B$_{2\mathrm{u}}$ magnetic structures without SOC. (b) The spin Berry curvature after taking band summation is integrated on the (010) planes for Col-B$_{3\mathrm{u}}$ (dark-blue), Col-A$_{\mathrm{u}}$ (brown), and Col-B$_{2\mathrm{u}}$ (dark-green) structure with SOC with its center changing from $-$Y to $\Gamma$, and to Y (illustrated as green rectangles in (a)). (c)-(e) The spin Berry curvature when the SOC is included for the Col-B$_{3\mathrm{u}}$, Col-A$_{\mathrm{u}}$, and Col-B$_{2\mathrm{u}}$ magnetic structures, respectively.}
\label{fig:kslice}
\end{figure}
Figure \ref{fig:bandshe}(b), (d), (f), and (h) show the SHC as a function of chemical potential for the non-magnetic, Col-B$_{3\mathrm{u}}$ magnetic, Col-A$_{\mathrm{u}}$ magnetic, and Col-B$_{2\mathrm{u}}$ magnetic structures. The result in the non-magnetic calculation shows an approximate antisymmetry for SHC components $\sigma^{\mathrm{S},\gamma}_{\alpha \beta}\approx -\sigma^{\mathrm{S},\gamma}_{\beta \alpha}$, while collinear-AFM states exhibit a large anisotropic SHE.
We illustrate an example for this anisotropy in Fig.~\ref{fig:illustrate}. 
Under electric fields along $x$ and $y$ direction, we have $\sigma^{\mathrm{S},z}_{xy}\approx -\sigma^{\mathrm{S},z}_{yx} $ for the non-magnetic state and $|\sigma^{\mathrm{S},z}_{xy}|\gg|\sigma^{\mathrm{S},z}_{yx}|$ for the Col-B$_{3\mathrm{u}}$ state. It implies that the spin current flowing along $x$-direction with $z$-spin component is significantly enhanced under an applied electric field along $y$-direction, ${\it i.e}$ $|j^{\mathrm{S},z}_{x,\mathrm{Col-B_{3u}}}|\gg| j^{\mathrm{S},z}_{x,\mathrm{non-mag}}|$, with the transition from non-magnetic state to AFM state by lowering temperature.
The drastic change of the SHC components by magnetic ordering might be useful for spintronic applications such as to control the spin current by varying the temperature. 

The largest component $\sigma^{\mathrm{S},z}_{xy}$ at zero energy have the magnitude 142.4 $(\hbar/e)$ S/cm, 182.0 $(\hbar/e)$ S/cm, and 140.8  $(\hbar/e)$ S/cm in the Col-B$_{3\mathrm{u}}$, Col-A$_{\mathrm{u}}$, and Col-B$_{2\mathrm{u}}$ magnetic structures, respectively. Each collinear-AFM state has a peak for $\sigma^{\mathrm{S},z}_{xy}$ located near the energy of Dirac points and nodal line, which is from 0 to about 30 meV below the Fermi energy. 
Strong dependence on chemical doping of the SHC is expected from the peak structure of chemical potential dependence of the SHC around Fermi level, whose maximum is $\sim$250 $(\hbar/e)$ S/cm, which is comparable with first-principles calculations results of ordinary metals such as Ta and W $\sim $200 $(\hbar/e)$ S/cm \cite{Tanaka_2008}.

Figure \ref{fig:kslice}(b) shows the spin Berry curvature after taking band summation, $\Omega^{\mathrm{S},z}_{xy,{\mathrm{sum}}}=\sum_{n}f_n \Omega^{\mathrm{S},z}_{n,xy}$, integrated over (010) planes with the center changing from $-$Y to $\Gamma$ and to Y in BZ. 
The plot shows that the largest contribution from the (010) plane including $\Gamma$ point and reduces with the plane center moving apart from $\Gamma$ along $y$-axis.
Figures~\ref{fig:kslice}(c)-(e) show the distribution of $\Omega_{xy,\mathrm{sum}}^{\mathrm{S},z}$, which contribute to the SHC $\sigma_{xy}^{\mathrm{S},z}$, in (010) plane containing the $\Gamma$ point within the first BZ, illustrated in Fig.~\ref{fig:kslice}(a).
The plot shows the large $\Omega_{xy}^{S,z}$
distributed around the gapped nodal line, implying the large contribution to the SHC from the Bloch states around the nodal line gapped out with the SOC in the (010) plane.
Distribution of the spin Berry curvature in $k$-space must reflect the symmetry constraint from the magnetic point group of the magnetic structure as discussed in Sec.~\ref{syms}.
As a result, under the mirror symmetry $m_y$=$PC_{2y}$ two-fold rotation $C_{2y}$, the symmetry of $\Omega_{xy,\mathrm{sum}}^{\mathrm{S},z}$ in Fig.~\ref{fig:kslice}(b) for Col-B$_{3\mathrm{u}}$ and Col-A$_{\mathrm{u}}$ structure, respectively  reflect as $\int\Omega_{xy,\mathrm{sum}}^{\mathrm{S},z}(k_x, -k_y, k_z)\mathrm{dS}=\int\Omega_{xy,\mathrm{sum}}^{\mathrm{S},z}(k_x, k_y, k_z)\mathrm{dS}$.
In Fig.~\ref{fig:kslice}(c) and Fig.~\ref{fig:kslice}(e) the spin Berry curvature of Col-B$_{3\mathrm{u}}$ and Col-B$_{2\mathrm{u}}$ magnetic structures hold the relation $\Omega_{xy,\mathrm{sum}}^{\mathrm{S},z}(k_x, k_y, -k_z)=\Omega_{xy,\mathrm{sum}}^{\mathrm{S},z}(k_x, k_y, k_z)$ with the mirror symmetry $m_z$=$PC_{2z}$.
\subsection{Spin Hall effect under external magnetic fields}
\begin{figure} 
\centering 
\includegraphics[width=7.0cm]{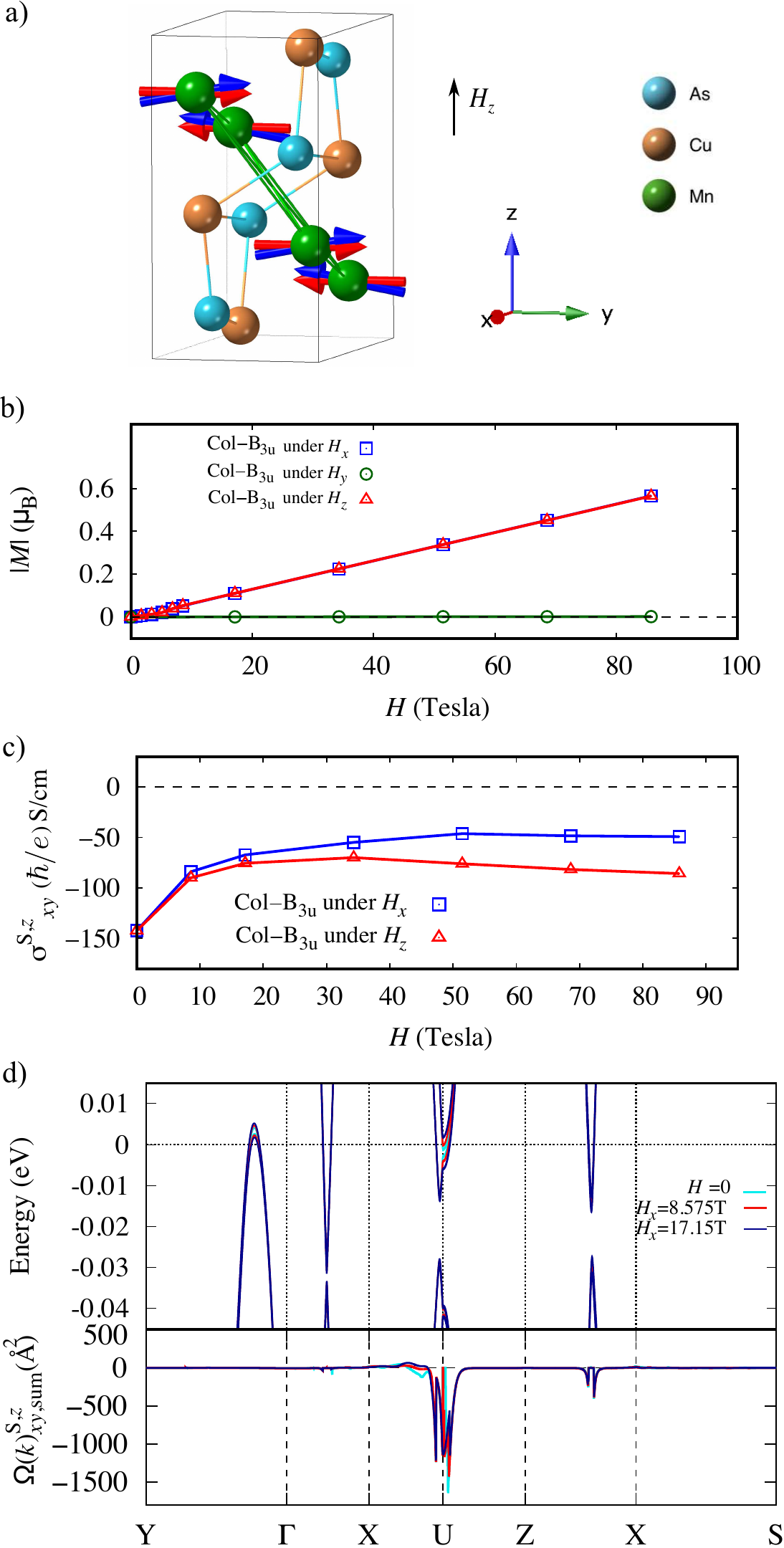}\\
\captionof{figure}{(a) Illustration changing of the local magnetic moment under $H_z$ for the Col-B$_{3\mathrm{u}}$ magnetic structure. The red vectors show magnetic moments in the ground state, the blue vectors show canted magnetic moments under $H_z$. (b) Magnetization under the magnetic fields $H_x, H_y, H_z$ in Col-B$_{3\mathrm{u}}$ magnetic structure. (c) The SHC $\sigma^{\mathrm{S},z}_{xy}$ under $H_x$ and $H_z$ in Col-B$_{3\mathrm{u}}$ magnetic structure. (d) The changing of electronic band structure and spin Berry curvature under the external magnetic fields along high-symmetry lines.}
\label{fig:magf}
\end{figure}
We consider Col-B$_{3\mathrm{u}}$ magnetic structure, which corresponds to the experimentally observed magnetic structure, under an applied magnetic fields along $x, y, z$-directions, $H_x$, $H_y$, and $H_z$. 
We illustrate the canting of the magnetic direction of Col-B$_{3\mathrm{u}}$ under the external magnetic fields along $z$-axis, $H_z$, in Fig.~\ref{fig:magf}(a). 
When the external magnetic fields are perpendicular to the direction of local magnetic moment, the absolute value of net magnetization increases with increasing the magnetic fields as shown in Fig.~\ref{fig:magf}(b) for the magnetic fields $H_x$ and $H_z$.
The external magnetic fields parallel to the local magnetic moments require a large field strength to flip the magnetic moments by conquering the AFM-exchange energy.
The large spin-flop transition from AFM to FM corresponding to the energy difference between the two magnetic ordered states shown in Table \ref{tab:energy} prevents magnetization from developing with the applied magnetic fields along $y$-axis, $H_y$ as seen in Fig.~\ref{fig:magf}(b). 
As a result, applying the magnetic field $H_y$ does not affect to the SHC in Col-B$_{3\mathrm{u}}$ structure within the investigated range of the magnetic fields in this work.

Figure \ref{fig:magf}(c) shows the SHC $\sigma^{\mathrm{S},z}_{xy}$ under the magnetic fields applied along $x$ and $z$-axis for Col-B$_{3\mathrm{u}}$ structure. The absolute values of the SHC reduces with increasing the external magnetic fields then it mostly reaches a saturation value when the external magnetic fields increases in the limit of considered applied magnetic fields from 0 to 85.75 Tesla.
We plot the electronic band structure and corresponding spin Berry curvature along the high-symmetry lines for the different magnitudes of external magnetic fields $H_x$ in Fig.~\ref{fig:magf}(d).
The magnetic fields strongly reduce the intensity of the spin Berry curvature component especially around the U point of the BZ, which the Dirac points are present around, against the small change in the band structure (See Fig.~\ref{fig:magf}(d)) and suppress the SHC as a result (See Fig.~\ref{fig:magf}(c)).
\subsection{Anomalous Hall effect under external magnetic fields}
\begin{figure} 
\centering 
\includegraphics[width=7.5cm]{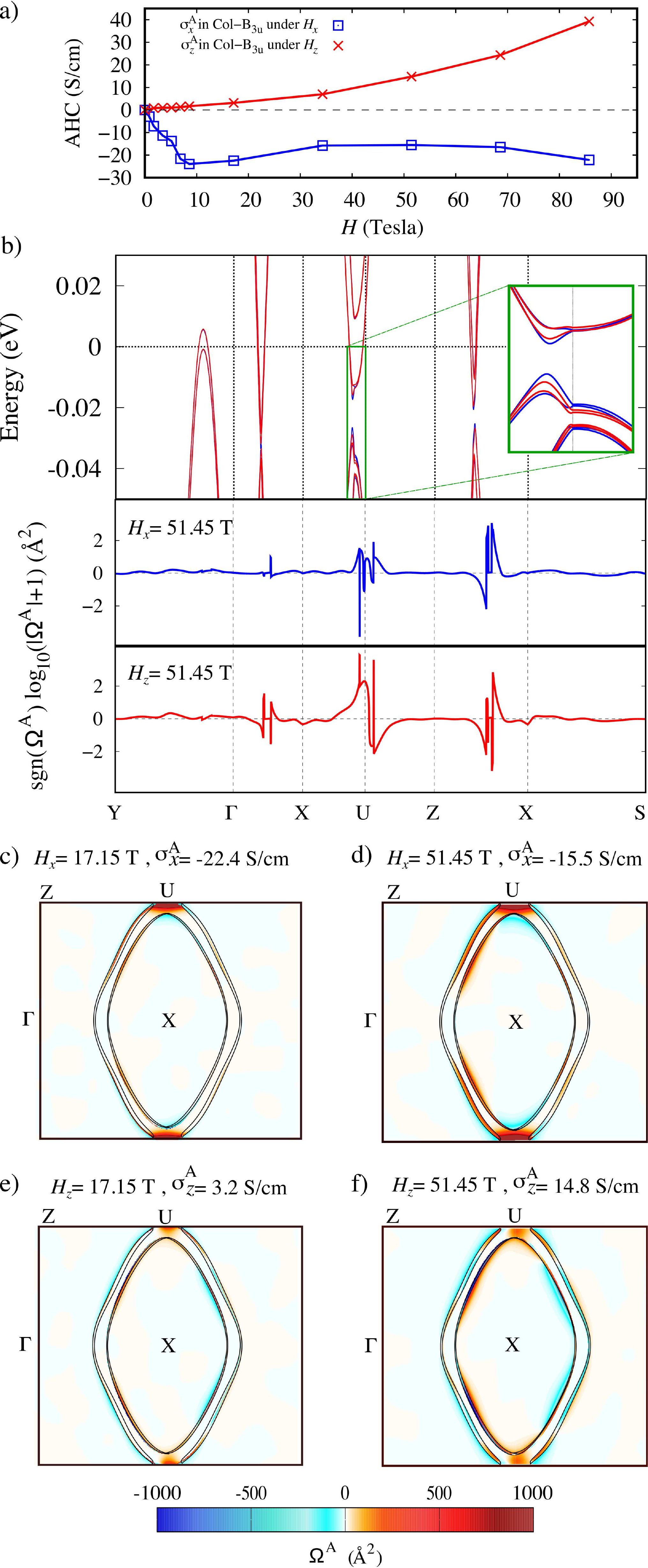}\\
\captionof{figure}{(a) The AHC $\sigma^\mathrm{A}_x=\sigma^\mathrm{A}_{yz}$ and $\sigma^\mathrm{A}_z=\sigma^\mathrm{A}_{xy}$ under $H_x$ and $H_z$ in Col-B$_{3\mathrm{u}}$ structure. (b) The electronic structure along high-symmetry lines, and the corresponding Berry curvature $\Omega^{\mathrm{A}}$ indicating $\Omega_x=\Omega^{\mathrm{A}}_{\mathrm{sum},yz}$ or $\Omega_z=\Omega^{\mathrm{A}}_{\mathrm{sum},xy}$ in the logarithm base 10 scale, $\mathrm{sgn}(\Omega^{\mathrm{A}})\log_{10}(|\Omega^{\mathrm{A}}|+1)$, under magnetic fields $H_x$ (blue) and $H_z$ (red) of 51.45 T. Here, sgn($x$) is the sign function, {\it i.e.} sgn$(x)=-1$ if $x<0$, sgn$(x)=0$ if $x=0$, and sgn$(x)=1$ if $x>0$. (c)-(f) Berry curvature in (010) plane containing $\Gamma$ for $H_x=\SI{17.15}{\tesla}$, $H_x=\SI{51.45}{\tesla}$, $H_z=\SI{17.15}{\tesla}$, $H_z=\SI{51.45}{\tesla}$, respectively, the black lines indicate Fermi surface.}
\label{fig:ahey}
\end{figure}
\begin{figure} 
\centering 
\includegraphics[width=7.5cm]{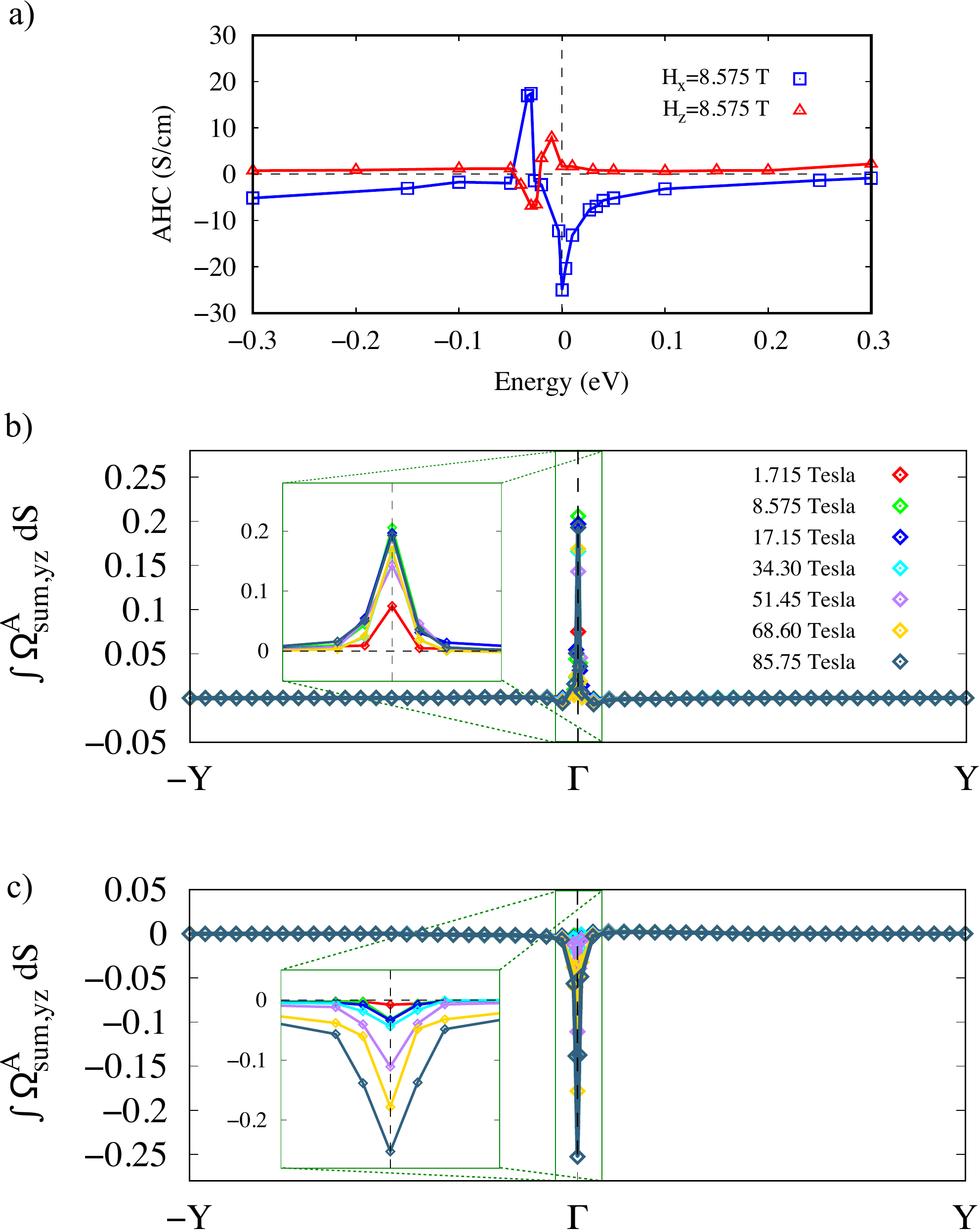}\\
\captionof{figure}{(a) The AHC as a function of the energy around Fermi energy in case of applied magnetic fields 8.575 T along $x$-axis and $z$-axis in Col-B$_{3\mathrm{u}}$ structure. The Berry curvature integrated on the (010) plane with its center changing from $-$Y to $\Gamma$, and to Y for Col-B$_{3\mathrm{u}}$ structure with SOC under the different (b) $H_x$ and (c) $H_z$.}
\label{fig:ahef}
\end{figure}
\begin{table}
\captionof{table}{The dependence of AHC on the $k$-mesh and adaptive mesh (Adapt.) at peaks of AHC values in Col-B$_{3\mathrm{u}}$ structure in Fig.~\ref{fig:ahef}(a).}
\label{tab:kmesh}
a) Under $H_x=\SI{8.575}{\tesla}$ 
\begin{tabular}{P{3.8cm}P{2.1cm}P{2.1cm}}
\hline
\hline
 & \multicolumn{2}{c}{ $\sigma^\mathrm{A}_x$ (\si[per-mode=symbol]{\siemens\per \centi\metre})}  \\
 \hline
$k$-mesh (Adapt.)  &  $E=\SI{-0.03}{\electronvolt}$ & $E=\SI{0}{\electronvolt}$ \\
\hline
 $180 \times 280 \times 180$  ($5 \times 5 \times 5$) & $16.9676$ & $-23.9518$ \\
$180 \times 280 \times 180$  ($7\times 7 \times 7$) & $17.7158$ &$ -23.9188$\\
 $200  \times  300  \times 200$  ($7 \times 7 \times 7$) & $17.4178$ & $-24.9715$ \\
\hline
\hline
\end{tabular}  \\
b) Under $H_z=\SI{8.575}{\tesla}$ 
\begin{tabular}{P{3.8cm}P{2.1cm}P{2.1cm}}
\hline
\hline
 & \multicolumn{2}{c}{ $\sigma^\mathrm{A}_z$ (\si[per-mode=symbol]{\siemens\per \centi\metre})}  \\
 \hline
$k$-mesh (Adapt.) &$E=\SI{-0.03}{\electronvolt}$& $
E=\SI{-0.01}{\electronvolt}$  \\
\hline
 $180 \times 280 \times 180$  ($5 \times 5 \times 5$) & $5.8248$ &$7.5697$\\
$180 \times 280 \times 180$  ($7\times 7 \times 7$) & $-6.6037$ &$7.4908$\\
 $200  \times  300  \times 200$  ($7 \times 7 \times 7$) & $-6.8678$ &$7.8558$\\
\hline
\hline
\end{tabular}  
\end{table}
In the collinear-AFM states of CuMnAs, the preserved $TP$ symmetry forbids the AHC being finite.
Meanwhile, we here show that the intrinsic AHC component can be developed significantly with the applied magnetic fields due to the local intensive contribution of the Berry curvature for the Bloch states related to the topological bands.
The result in Fig.~\ref{fig:ahey}(a) shows the development of the AHC under the different directions of the applied magnetic fields $H_x$ and $H_z$ which cause the linear development of the magnetization in Col-B$_{3\mathrm{u}}$ structure in Fig.~\ref{fig:magf}(b). Under the magnetic field along $x$-axis, the magnitude of AHC $\sigma^{\mathrm{A}}_{x}$  increases from zero to 27 S/cm at $H_x$ about 8.575 Tesla and shows only a weak change above the fields. 
Meanwhile, the AHC $\sigma^{\mathrm{A}}_{z}$ increases monotonically
with increasing magnetic field along the $z$-axis. 

We show the electronic band structure and corresponding Berry curvature along the high-symmetry lines in case of the applied magnetic fields 51.45 Tesla along $x$ and $z$-axis in Fig.~\ref{fig:ahey}(b). As shown in Fig.~\ref{fig:ahey}(b), the Berry curvature around U
point exhibits a large difference for the different directions of applied magnetic fields against the little change in the band structure.
The difference can be seen more clearly for distribution of the Berry curvature in (010) plane including the $\Gamma$ point for different directions as plotted in Fig.~\ref{fig:ahey}(c)-(f). 
The figure shows the stronger intensity of the Berry curvature around U point under $H_x$ magnetic fields than that under $H_z$ fields while the small Berry curvature under $H_z$ fields spread more around the gapped nodal lines than that under $H_x$ fields.
Since the local peaky Berry curvature can change drastically for the small change in the energy bands and widely spread small Berry curvature provides more gradual change in the contribution to the AHC, the difference in the development of AHC for different magnetic fields direction can be explained by the different origin of the Berry curvature, as shown in  Fig.~\ref{fig:ahey}(c)-(f). 

We show the AHC value as a function of chemical potential in Col-B$_{3\mathrm{u}}$ structure at the external magnetic field 8.575 Tesla in Fig.~\ref{fig:ahef}(a).
The result shows the drastic change of magnitude and sign of AHC values around the energy of the gapped out nodal line, which is about from 0 to 30 meV below the Fermi energy.
The rapid change implies the possible change in AHC value via chemical doping and partly help us to realize origin of the AHC possible from the gapped nodal line. We increase the $k$-mesh to check the convergence around the region with rapid change shown in Table \ref{tab:kmesh}. 
To see where the largest contribution of Berry curvature to the AHC is, we estimated the sum of Berry curvature in (010) planes with its center changing from $-$Y to $\Gamma$ and to Y for the different intensity values of $H_x$ and $H_z$ in Col-B$_{3\mathrm{u}}$ structure. 
The result in Fig.~\ref{fig:ahef}(b) and Fig.~\ref{fig:ahef}(c) shows that the largest summation belongs to the (010) plane including $\Gamma$ which contains the gapped out Dirac nodal line by SOC. 
The microscopic mechanism leading to significant anomalous components of the Hall effect predicted for the AFM CuMnAs is therefore concluded to be the further lifting of band degeneracy under external magnetic fields for the Bloch states generated with splitting of nodal lines by SOC near Fermi energy.
\section{Conclusions}\label{conclusion}
In summary, we investigated the influence of topological bands in macroscopic transport phenomena, SHE and AHE, in orthorhombic CuMnAs. 
We provided the symmetry analysis for these transport phenomena and magnetic structures in the crystal structure. 
The collinear-AFM states in CuMnAs form Dirac nodal-lines in the energy bands near the Fermi energy in the absence of SOC and the nodal lines generate a large SHC with lifting degeneracies by SOC.
The change of the SHC from the asymmetric tensor components in the nonmagnetic state to anisotropic ones in the AFM states enables us manipulate the spin current by controlling magnetism and is expected to be applicable as spintronics devices~\cite{Zhang,turnable,Zelezny_2017,Brink_2018,Kimata_2019,Naka_2019,Ahn_2019,Hayami_2019,Hayami_2020_ncm,Hayami_2020_full,Mook_2020,Yuan_2020,Naka_2021,Hernandez_2021}. 

An experimentally observed collinear-AFM state referred as Col-B$_{3\mathrm{u}}$ structure, in which the local magnetic moments aligned along the $b$-axis, is investigated with the effect of applied magnetic fields.
Applied magnetic fields cause drastic change in the SHC and AHC with the local contribution from the Bloch states around the nodal lines opening the gap with SOC through further lifting of the band degeneracy due to $TP$ symmetry breaking, against the simple linear development of the magnetization.

The rapid chemical potential dependence in the SHC and AHC leads to possibility to manipulate these transport phenomena via chemical doping in CuMnAs. The SHC and AHC in the collinear AFM state thus can pave a way to control these phenomena in the fields of spintronics as well as to investigate the transport phenomena for further AFM materials.
\section*{Acknowledgement}
 We would like to thank K.~K.~Huynh and M.~Kimata for fruitful discussions of experimental results on CuMnAs. One of the present authors (V.~T.~N.~Huyen) also thanks O.~Gr\aa n\"as for useful discussion related to using the ELK code. This research is supported by JSPS KAKENHI Grants Numbers JP19H01842, JP20H05262, JP20K05299, JP20K21067, JP21H01789, JP21H04437, JP21H01031, and by JST PRESTO Grant Number JPMJPR17N8. We also acknowledge the use of supercomputing system, MASAMUNE-IMR, at CCMS, IMR, Tohoku University in Japan.

\bibliography{basename of .bib file}

\begin{thebibliography}{99}
\bibitem{nagaosa} N. Nagaosa, J. Sinova, S. Onoda, A. H. MacDonald, and N. P. Ong, Rev. Mod. Phys. {\bf 82}, 1539 (2010).
\bibitem{xiao} D. Xiao, M. C. Chang, and Q. Niu, Rev. Mod.  Phys. {\bf 82}, 1959 (2010).
\bibitem{Baltz_2018} V. Baltz, A. Manchon, M. Tsoi, T. Moriyama, T. Ono, and Y. Tserkovnyak, Rev. Mod. Phys. \textbf{90}, 015005 (2018). 
\bibitem{2015mn3sn} S. Nakatsuji, N. Kiyohara, and T. Higo, Nature {\bf 527}, 212 (2015).
\bibitem{Kiyohara_2016} N. Kiyohara, T. Tomita, and S. Nakatsuji, Phys. Rev. Applied {\bf 5}, 064009 (2016).
\bibitem{2014Chen} H. Chen, Q. Niu, and A. H. MacDonald, Phys. Rev. Lett. {\bf 112}, 017205 (2014).
\bibitem{2014mn3sn} J. K\"ubler and C. Felser,  Europhys. Lett. {\bf 108}, 67001 (2014).
\bibitem{mn3an} V. T. N. Huyen, M.-T. Suzuki, K. Yamauchi, and T. Oguchi, Phys. Rev. B {\bf 100}, 094426 (2019).
\bibitem{Sakai_co2mnga_2018} A. Sakai, Y. P. Mizuta, A. A. Nugroho, R. Sihombing, T. Koretsune, M.-T. Suzuki, N. Takemori, R. Ishii, D. Nishio-Hamane, R. Arita, P. Goswami, and S. Nakatsuji, Nat. Phys. \textbf{14}, 1119 (2018).
\bibitem{Guin_co3sn2s2_2019} S. N. Guin, P. Vir, Y. Zhang, N. Kumar, S. J. Watzman, C. Fu, E. Liu, K. Manna, W. Schnelle, J. Gooth, C. Shekhar, Y. Sun, and C. Felser, Adv. Mater. \textbf{31}, 1806622 (2019).
\bibitem{Minami_2020} S. Minami, F. Ishii, M. Hirayama, T. Nomoto, T. Koretsune, and R. Arita, Phys. Rev. B \textbf{102}, 205128 (2020). 
\bibitem{Yanagi_2020} Y. Yanagi, J. Ikeda, K. Fujiwara, K. Nomura, A. Tsukazaki, and M.-T. Suzuki, arXiv:2011.14567 (2020). 
\bibitem{Tanaka_2008} T. Tanaka, H. Kontani, M. Naito, T. Naito, D. S. Hirashima, K. Yamada, and J. Inoue, Phys. Rev. B \textbf{77}, 165117 (2008).
\bibitem{Freimuth} F. Freimuth, S. Bl\"ugel, and Y. Mokrousov, Phys. Rev. Lett. {\bf 105}, 246602 (2010).
\bibitem{Zhang} W. Zhang, M. B. Jungfleisch, W. Jiang, J. E. Pearson, and A. Hoffmann, F. Freimuth, and Y. Mokrousov, Phys. Rev. Lett. {\bf 113}, 196602 (2014).
\bibitem{turnable} Y. Yen and G.-Y. Guo, Phys. Rev. B {\bf 101}, 064430 (2020).
\bibitem{cumnas3} P. Tang, Q. Zhou, G. Xu, and S.-C. Zhang, Nat. Phys. {\bf 12}, 1100 (2016).
\bibitem{cumnas4} L. \v{S}mejkal, J. \v{Z}elezn\'{y}, J. Sinova, and T. Jungwirth, Phys. Rev. Lett. {\bf 118}, 106402 (2017).
\bibitem{cumnas5} F. M\'aca, J. Ma\v sek, O. Stelmakhovych, X. Mart\'i, H. Reichlov\'a, K. Uhl\'i\v rov\'a, P. Beran, P. Wadley, V. Nov\'ak, and T. Jungwirth, J. Magn. Magn. Mater. {\bf 324}, 1606 (2012).
\bibitem{cumnas6} E. Emmanouilidou, H. Cao, P. Tang, X. Gui, C. Hu, B. Shen, J. Wu, S.-C. Zhang, W. Xie, and N. Ni, Phys. Rev. B {\bf 96}, 224405 (2017).
\bibitem{berry1984} M. V. Berry, Proc. R. Soc. Lond. A {\bf 392}, 45 (1984).
\bibitem{velocity3} F. D. M. Haldane, Phys. Rev. Lett. {\bf 93}, 206602 (2004).
\bibitem{ahc4} D. J. Thouless, M. Kohmoto, M. P. Nightingale, and M. den Nijs, Phys. Rev. Lett. {\bf 49}, 405 (1982).
\bibitem{ahc} X. Wang, J. R. Yates, I. Souza, and D. Vanderbilt, Phys. Rev. B {\bf 74}, 195118 (2006).
\bibitem{QE} P. Giannozzi, S. Baroni, N. Bonini, M. Calandra, R. Car, C. Cavazzoni, D. Ceresoli, G. L. Chiarotti, M. Cococcioni, I. Dabo,  A. D. Corso, S. de Gironcoli, S. Fabris, G. Fratesi, R. Gebauer, U. Gerstmann, C. Gougoussis, A. Kokalj, M. Lazzeri, L. M.-Samos, N. Marzari, F. Mauri, R. Mazzarello, S. Paolini, A. Pasquarello, L. Paulatto, C. Sbraccia, S. Scandolo, G. Sclauzero, A. P Seitsonen, A. Smogunov, P. Umari and R. M Wentzcovitch, J. Phys. Condens. Matter {\bf 21}, 395502 (2009).
\bibitem{GGA-PBE} J. P. Perdew, K. Burke, and M. Ernzerhof, Phys. Rev. Lett. {\bf 77}, 3865 (1996).
\bibitem{paw_original} P. E. Bl\"{o}chl, Phys. Rev. B \textbf{50}, 17953 (1994).
\bibitem{paw} G. Kresse and D. Joubert, Phys. Rev. B {\bf 59}, 1758 (1999).
\bibitem{pslibrary} A. D. Corso, Comput. Mater. Sci., {\bf 95}, 337 (2014).
\bibitem{paoflow} M. B. Nardelli, F. T. Cerasoli, M. Costa, S Curtarolo, R. De Gennaro, M. Fornari, L. Liyanage, A. R. Supka and H. Wang, Comput. Mater. Sci. {\bf 143}, 462 (2018).
\bibitem{elk} J. K. Dewhurst {\it et al.}, \url{http://elk.sourceforge.net/}.
\bibitem{w90} A. A. Mostofi, J. R. Yates, G. Pizzi, Y. S. Lee, I. Souza, D. Vanderbilt, and N. Marzari, Comput. Phys. Commun. {\bf 185}, 2309 (2014).
\bibitem{adap1}Y. Yao, L. Kleinman, A. H. MacDonald, J. Sinova, T. Jungwirth, D.-S. Wang, E. Wang, and Q. Niu, Phys. Rev. Lett. {\bf 92}, 037204 (2004).
\bibitem{adap2} N. Marzari and D. Vanderbilt, Phys. Rev. B {\bf 56}, 12847 (1997).
\bibitem{cmpgeneration} M.-T. Suzuki, T. Nomoto, R. Arita, Y. Yanagi, S. Hayami, H. Kusunose, Phys. Rev. B {\bf 99}, 174407 (2019).
\bibitem{Grimmer_1993} H. Grimmer, Acta Crystallogr. Sect. A \textbf{49}, 763 (1993).
\bibitem{Martinez_2015} D. Gos\'{a}lbez-Mart\'{i}nez, I. Souza, and D. Vanderbilt, Phys. Rev. B \textbf{92}, 085138 (2015).  
\bibitem{cmp2017} M.-T. Suzuki, T. Koretsune, M. Ochi, and R. Arita, Phys. Rev. B {\bf 95}, 094406 (2017).
\bibitem{seemann} M. Seemann, D. K\"odderitzsch, S. Wimmer, and H. Ebert, Phys. Rev. B {\bf 92}, 155138 (2015).
\bibitem{Huebsch} M.-T. Huebsch, T. Nomoto, M.-T. Suzuki, and R. Arita, Phys. Rev. X {\bf 11}, 011031 (2021).
\bibitem{accuracy_gga} L. He, F. Liu, G. Hautier, M. J. T. Oliveira, M. A. L. Marques, F. D. Vila, J. J. Rehr, G.-M. Rignanese, and A. Zhou, Phys. Rev. B {\bf 89}, 064305 (2014).
\bibitem{Zelezny_2017} J. \ifmmode \check{Z}\else \v{Z}\fi{}elezn\'y, Y. Zhang, C. Felser, and B. Yan, Phys. Rev. Lett. \textbf{119}, 187204 (2017).
\bibitem{Brink_2018} Y. Zhang, J. \v{Z}elezn\'{y}, Y. Sun, J. van den Brink, and B. Yan, New J. Phys. \textbf{20}, 073028 (2018).
\bibitem{Kimata_2019} M. Kimata, H. Chen, K. Kondou, S. Sugimoto, P. K. Muduli, M. Ikhlas, Y. Omori, T. Tomita, A. H. MacDonald, S. Nakatsuji, and Y. Otani, Nature (London) \textbf{565}, 627 (2019).
\bibitem{Naka_2019} M. Naka, S. Hayami, H. Kusunose, Y. Yanagi, Y. Motome, and H. Seo, Nat. Commun. \textbf{10}, 4305 (2019).
\bibitem{Ahn_2019} K.-H. Ahn, A. Hariki, K.-W. Lee, and J. Kune\v{s}, Phys. Rev. B \textbf{99}, 184432 (2019).
\bibitem{Hayami_2019} S. Hayami, Y. Yanagi, and H. Kusunose, J. Phys. Soc. Jpn. \textbf{88}, 123702 (2019).
\bibitem{Hayami_2020_ncm}  S. Hayami, Y. Yanagi, and H. Kusunose, Phys. Rev. B \textbf{101}, 220403(R) (2020).
\bibitem{Hayami_2020_full} S. Hayami, Y. Yanagi, and H. Kusunose, Phys. Rev. B \textbf{102}, 144441 (2020).
\bibitem{Mook_2020} A. Mook, R. R. Neumann, A. Johansson, J. Henk, and I. Mertig, Phys. Rev. Research \textbf{2}, 023065 (2020).
\bibitem{Yuan_2020} L.-D. Yuan, Z. Wang, J.-W. Luo, E. I. Rashba, and A. Zunger, Phys. Rev. B \textbf{102}, 014422 (2020).
\bibitem{Naka_2021} M. Naka, Y. Motome, and H. Seo, Phys. Rev. B \textbf{103}, 125114 (2021).
\bibitem{Hernandez_2021} R. Gonz\'alez-Hern\'andez, L. \ifmmode \check{S}\else \v{S}\fi{}mejkal, K. V\'yborn\'y, Y. Yahagi, J. Sinova, T. Jungwirth, and J. \ifmmode \check{Z}\else \v{Z}\fi{}elezn\'y, Phys. Rev. Lett. \textbf{126}, 127701 (2021). 
%
\end{thebibliography}

\end{document}